\documentclass[3p]{elsarticle}
\usepackage{amsmath}

\usepackage{array}
\usepackage{graphicx}

\usepackage{subfigure}
\usepackage{subfigmat}

\usepackage{xcolor}

\usepackage{setspace}

\usepackage{tikz}
\usetikzlibrary{shapes,arrows}

\usepackage{float}
\begin{document}

\begin{frontmatter}

\title{Combustion Instability of a Multi-injector Rocket Engine Using the Flamelet Progress Variable Model}

\author[rvt]{Lei~Zhan}
\author[focal]{Tuan~M.~Nguyen}
\author[rvt]{Juntao~Xiong}
\author[rvt]{Feng~Liu}
\author[rvt]{William~A.~Sirignano}

\address[rvt]{Department of Mechanical and Aerospace Engineering, University of California, Irvine, California 92697, USA}
\address[focal]{Sandia National Laboratories, Livermore, CA 94550, USA}

\begin{abstract}

The combustion instability is investigated computationally for a multi-injector rocket engine using the flamelet progress variable (FPV) model. An C++ code is developed based on OpenFOAM 4.0 to apply the combustion model. Flamelet tables are generated for methane/oxygen combustion at the background pressure of $200$ bar using a 12-species chemical mechanism. A power law is determined for rescaling the reaction rate for the progress variable to address the pressure effect. The combustion is also simulated by the one-step-kinetics (OSK) method for comparison with the FPV approach. Study of combustion instability shows that a longitudinal mode of $1500$ Hz and a tangential standing wave of $2500$ Hz are dominant for both approaches. While the amplitude of the longitudinal mode remains almost the same for both approaches, the tangential standing wave achieves larger amplitude in the FPV simulation. A preliminary study of the resonance in the injectors, which is driven by the longitudinal-mode oscillation in the combustion chamber, is also presented.

\end{abstract}

\end{frontmatter}

\section{Introduction}

Combustion instability is an acoustical phenomenon in which an unstable pressure oscillation is  excited and sustained by combustion. In most high-power propulsion systems that propel rockets and airplanes, the high-energy release rate reinforces acoustical oscillations with very high amplitudes. Such oscillations can affect thrust in an undesirable way, and sometimes lead to engine destruction. 

In recent years, there is an increasing need for three-dimensional high-fidelity numerical study of combustion instability in multi-injector rocket engines. Compared to the single-injector configuration, the combustion instability mechanism is more complex in a multi-injector combustion chamber because of the possible interactions among injectors and interactions with the complicated feed system and upstream manifold. Systematic experimental investigations of combustion instability for such configurations are quite expensive and technically difficult. Compared to experimental measurement, numerical simulations based on high-fidelity turbulence modeling and detailed chemical reaction mechanisms are able to provide more details about the flow physics as well as the combustion dynamics and hence gain a deeper insight into the combustion instability mechanisms. Urbano et al. \cite{multicnf} carried out combustion simulations for a complete small-scale rocket engine with 42 coaxial injectors and a nozzle outlet. Large eddy simulation (LES) was employed and a four-species $H_{2}$/$O_{2}$ reacting mechanism was considered. Combustion instability was studied with emphasis on interaction between transverse and radial modes. Xiong et al. \cite{xiongaiaaj2020} presented computational study of the nonlinear combustion instability of a similar configuration, which has a choked nozzle and a combustion chamber fed by 10 or 19 coaxial injector ports. Detailed time-resolved information about the combustion instability was provided using a three-dimensional unsteady $k-\omega$ shear-stress transport delayed detached-eddy simulation. The global one-step $CH_{4}$/$O_{2}$ reaction was adopted. The triggered tangential and spontaneous as well as triggered longitudinal instability modes were explored.

While high-fidelity turbulence modeling were commonly adopted, the combustion models in many studies are limited to either one- or two-step chemical mechanisms involving only four to five species. In Ref.\cite{xiongaiaaj2020}, the filtered/Favre-averaged species transport equations were solved and the laminar closure model (LCM) was used to handle the filtered reaction source terms. With this finite-rate chemistry model, it is difficult to incorporate any realistic detailed-reaction mechanism involving many species and reactions due to huge computational cost. Additionally, the stiffness arising from nonlinearity of species reaction source terms and the wide range of chemical time scales makes the system very difficult to solve \cite{tuancnf}. The transported probability density function (PDF) approach \cite{popcnf} is another popular option. Although any additional model for the chemical source terms is not required, the PDF simulations are usually very computationally expensive even with a simple chemistry mechanism due to the time-consuming solving process \cite{haworthpecs}.

As an alternative method, the flamelet approach is based on the assumption that the chemical time scales are shorter than the turbulent time scales. Hence, flame in the turbulent scenario is in local equilibrium and can be viewed as a collection of laminar flamelets, which are one-dimensional and only have structures in the direction perpendicular to the surface of the stoichiometric mixture \cite{nppecs, nptc}. The definition of flamelet allows an independent chemistry computation prior to the main flow simulation and the flame solutions can be prepared in the flamelet
libraries/tables. Therefore, detailed chemical mechanisms can be incorporated without bringing additional computational cost on the main flow calculations. For non-premixed turbulent combustion, a successful method is the laminar diffusion flamelet approach \cite{nppecs}. It is called the steady laminar flamelet (SLF) model in this report as the steady flamelet equations are solved in the mixture fraction space. In the SLF model, scalar dissipation rate is used to parameterize the flamelet tables and each scalar dissipation rate could be related to multiple solutions. Therefore, despite the high computational efficiency, the SLF model only explores a part of the flamelet solutions so that the selected solutions are single-valued functions of the scalar dissipation rate. To include all the flamelet solutions in the flamelet approach, Pierce and Moin \cite{piercemoinjfm} introduced the flamelet progress variable (FPV) approach. The scalar dissipation rate is replaced by the reaction progress variable, which is defined as the total mass fraction of selected major species in a chemical mechanism, in parameterization of the flamelet tables. This transition makes the flamelet solutions well-defined functions of all parameters including the progress variable. The baseline FPV approach were primarily applied in simulations of incompressible combustion. In recent years, the use of the FPV approach was extended to the computations of compressible combustion by considering the kinetic energy in the energy balance regime \cite{pecnikaiaaj2012,saghafiancnf}. Tuan et al. \cite{tuanjpp} examined the capability of the PFV model in simulating combustion instability in the continuously variable resonance chamber of a single-injector rocket engine. The numerical results for pressure oscillation frequencies and amplitudes showed good agreement with the experiment data.

In the current research, our goal is to perform three-dimensional high-fidelity turbulent combustion simulation of a ten-injector rocket engine and conduct combustion instability study using the the efficient FPV approach incorporating detailed chemical reaction mechanisms. In this report, we present our progress in developing the OpenFOAM-based code and the primary computational results.

\section{The Flamelet Progress Variable Approach}

\subsection{The Governing Equations}
\label{sec:governequ}

For a mixture of multi-species, the Favre-averaged Navier-Stokes equations are written in the following conservative form as 

\begin{equation}
\frac{\partial \bar{\rho}}{\partial t} + \frac{\partial \bar{\rho} \widetilde v_{j} }{\partial x_{j}}  = 0
\label{continuityequation}
\end{equation}

\begin{equation}
\frac{\partial \bar{\rho} \widetilde v_{i}}{\partial t} + \frac{\partial \bar{\rho} \widetilde v_{i} \widetilde v_{j}}{\partial x_{j}} = -\frac{\partial{\bar{p}}}{\partial x_{i}} + \frac{\partial(\tau_{i,j}+\tau^{R}_{i,j})}{\partial x_{j}}
\label{momentumequation}
\end{equation}

\begin{equation}
\frac{\partial \bar{\rho} \widetilde{h_{a}}}{\partial t} + \frac{\partial \bar{\rho} \widetilde v_{j} \widetilde{h_{a}}}{\partial x_{j}} + \frac{\partial \bar{\rho} \widetilde{K}}{\partial t} + \frac{\partial \bar{\rho} \widetilde v_{j} \widetilde{K}}{\partial x_{j}} - \frac{\partial \widetilde{p}}{\partial t} = \frac{\partial [\widetilde v_{i}(\tau_{i,j}+\tau^{R}_{i,j})]}{\partial x_{j}} + \frac{\partial}{\partial x_{j}}[(\frac{\lambda}{c_{p}}+\frac{\mu_{t}}{Pr_{t}})\frac{\partial \widetilde{h_{a}}}{\partial x_{j}}]
\label{haequation}
\end{equation}
where $\bar{\rho}$ is the mean density, $\widetilde v_{j}$ is the mean velocity vector and $\bar{p}$ is the mean pressure. In the energy equation, $\widetilde{h_{a}}$ is the mean total(or absolute) enthalpy, which is the summation of the mean sensible enthalpy $\widetilde{h_{s}}$ and the mean enthalpy of formation $\widetilde{h_{c}}$, and $\widetilde{K} = \frac{1}{2}(\sum^{n}_{j=1}\widetilde v_{j}\widetilde v_{j})$ is the mean kinetic energy. $\lambda$ and $C_{p}$ are the heat conduction coefficient and the specific heat at constant pressure. $Pr_{t}$ is turbulent Prandtl number. $\tau_{i,j}$ and $\tau^{R}_{i,j}$ are the molecular and turbulent viscous stress tensors and expressed as 
$\tau_{i,j} = \mu(\frac{\partial \widetilde v_{i}}{\partial x_{j}}+\frac{\partial \widetilde v_{j}}{\partial x_{i}}-\frac{2}{3}\frac{\partial \widetilde v_{k}}{\partial x_{k}}\delta_{i,j})$
and 
$\tau^{R}_{i,j} = \mu_{t}(\frac{\partial \widetilde v_{i}}{\partial x_{j}}+\frac{\partial \widetilde v_{j}}{\partial x_{i}}-\frac{2}{3}\frac{\partial \widetilde v_{k}}{\partial x_{k}}\delta_{i,j})$
with $\mu$ and $\mu_{t}$ being the molecular and turbulent viscosity.

\subsection{The Turbulent Combustion Model}
\label{sec:turcombusmodel}

In computations of turbulent combustion based on the Reynolds-Averaged Navier-Stokes (RANS) methods, such as those in this report, only the mean quantities are transported. Hence, the laminar flamelet solutions need to be convoluted with assumed probability density functions of the independent scalars, including mean mixture fraction $\widetilde{Z}$ and the mean variance $\widetilde{Z^{''2}}$. The convolution generates flamelet libraries of mean quantities which can be accessed efficiently during the computational fluid dynamics (CFD). In the FPV variant of the flamelet model, the flamelet tables are also parameterized by a mean reaction progress variable $\widetilde{C}$. In this report, the progress variable is defined as the total mass of four major species, which are $CO_{2}$, $H_{2}O$, $H_{2}$ and $CO$. Hence, a mean thermal and chemical quantities $\widetilde{\psi}_{i}$, such as composition and temperature, is given by

\begin{equation}
\widetilde{\psi}_{i}(\widetilde{Z},\widetilde{Z^{''2}},\widetilde{C}) = \int^{1}_{0}\int^{1}_{0}\psi_{i}(Z,C)\tilde{P}(Z,Z^{''2},C)dZdC
\label{convolution}
\end{equation}
where $\tilde{P}$ is the PDF function for $\widetilde{\psi}_{i}$. The PDF for the progress variable is frequently assumed to be the Dirac $\delta$ function, thus, \eqref{convolution} is reduced to 
\begin{equation}
\widetilde{\psi}_{i}(\widetilde{Z},\widetilde{Z^{''2}},\widetilde{C}) = \int^{1}_{0}\int^{1}_{0}\psi_{i}(Z,\widetilde{C})\tilde{P}(Z,Z^{''2})dZ
\label{reduceconvolution}
\end{equation}
Finally, the $\beta$ PDF is assumed for $\tilde{P}$, which gives
\begin{equation}
\tilde{P}(Z,Z^{''2} = \frac{Z^{\alpha-1}(1-Z)^{\beta-1}}{\Gamma(\alpha)\Gamma(\beta)}\Gamma(\alpha+\beta)
\label{betapdf}
\end{equation}
where $\Gamma$ is the gamma function. $\alpha$ and $\beta$ are shape functions which are defined as $\alpha = Z\gamma$ and $\beta = (1-Z)\gamma$ with $\gamma = [Z(1-Z)/Z^{''2}]-1 \ge 0$. 
  
For compressible combustion, the tables should be generated at different background pressures and pressure should be included as an input to access the tables. Pecnik et al. \cite{pecnikaiaaj2012} showed that the composition of the reacting mixture is not sensitive to the  background pressure and only the reaction rate changes dramatically. This fact justifies the method of using tables under one background pressure (usually the averaged pressure) and rescaling the reaction rate for other pressures. In this report, we adopt this approach to simplify preparation of the tables and the table lookup process.

At each time step in the CFD computation, the transport equations for scalars $\widetilde{Z}$, $\widetilde{Z^{''2}}$ and $\widetilde{C}$ are solved. When the Levis number is equal to one, these equations are given as  

\begin{equation}
\frac{\partial \bar{\rho} \widetilde{Z}}{\partial t} + \frac{\partial \bar{\rho} \widetilde v_{j} \widetilde{Z}}{\partial x_{j}} = \frac{\partial}{\partial x_{j}}[(\frac{\mu}{Sc}+\frac{\mu_{t}}{Sc_{t}})\frac{\partial \widetilde{Z}}{\partial x_{j}}]
\label{zequation}
\end{equation}

\begin{equation}
\frac{\partial \bar{\rho} \widetilde{Z^{''2}}}{\partial t} + \frac{\partial \bar{\rho} \widetilde v_{j} \widetilde{Z^{''2}}}{\partial x_{j}} = \frac{\partial}{\partial x_{j}}[(\frac{\mu}{Sc}+\frac{\mu_{t}}{Sc_{t}})\frac{\partial \widetilde{Z^{''2}}}{\partial x_{j}}]+
2\frac{\mu_{t}}{Sc_{t}}\frac{\partial \widetilde{Z}}{\partial x_{j}}\frac{\partial \widetilde{Z}}{\partial x_{j}}-\bar{\rho}\widetilde{\chi}
\label{varzequation}
\end{equation}

\begin{equation}
\frac{\partial \bar{\rho} \widetilde{C}}{\partial t} + \frac{\partial \bar{\rho} \widetilde v_{j} \widetilde{C}}{\partial x_{j}} = \frac{\partial}{\partial x_{j}}[(\frac{\mu}{Sc}+\frac{\mu_{t}}{Sc_{t}})\frac{\partial \widetilde{C}}{\partial x_{j}}]+\widetilde{\dot{\omega}_{C}}
\label{cequation}
\end{equation}
where $\widetilde \chi$ is the Favre-averaged scalar dissipation rate. By comparing the integral scalar time scale and the turbulent flow time scale, $\widetilde \chi$ is frequently expressed as $\widetilde \chi = C_{\chi}\frac{\tilde \epsilon}{\tilde k}\widetilde{Z^{''2}}$ with $\tilde k$ and $\tilde \epsilon$ being the mean turbulent kinetic energy and mean turbulent dissipation. A constant value of $2.0$ is often used for $C_{\chi}$. The laminar and turbulent Schmidt number $Sc$ and $Sc_{t}$ are $1.0$ and $0.7$, respectively. The source term $\widetilde{\dot{\omega}_{C}}$ in equation \eqref{cequation} is the reaction rate for the progress variable.

\subsection{The laminar flame calculator}
\label{sec:flamemasterffcmy12}

In this report, the laminar flamelet solutions are generated using FlameMaster, an open source C++ program package for zero-dimensional combustion and one-dimensional laminar flame calculations\cite{flamemaster}. The code can be used for steady and unsteady computations of premixed as well as non-premixed flames. For the methane/oxygen combustion that we consider in this report, we use the module of the code that solves the steady flamelet equations in the mixture fraction space by assuming counterflow configuration. To run the code, a chemical mechanism and corresponding thermo and transport data are needed. The background pressure, composition and temperature on both the fuel and oxidizer sides are required. For each given stoichiometric scalar dissipation rate ($\chi_{st}$), the code generates a flame solution in form of distributions of temperature, composition, reaction rates and heat release rate in the mixture fraction space. Hence the laminar flamelet solutions are already parameterized by the scalar dissipation rate and the mixture fraction.

Since chamber pressure in the current study is around $200$ bar, a chemical mechanism suitable 
for high-pressure combustion is desired. We adopt a skeletal model of 12 species, which is 
generated by selecting relevant species and reaction pathways from version y of a 119-species 
Foundation Fuel Chemistry Model (FFCM-y)\cite{stanfordffcmy,ffcmcnf}.
The FFCM models were recently developed with uncertainty minimization against up-to-date 
experimental data for the combustion of small hydrocarbon fuels. The models were tested for 
fuels, such as methane, over the pressure range of $10$-$100$ atm. The 12 species in the skeletal model are $H_{2}$,  
$H$, $O_{2}$, $O$, $OH$, $HO_{2}$, $H_{2}O$, $CH_{3}$, $CH_{4}$, $CO$, $CO_{2}$, and 
$CH_{2}O$. Reaction rate constants in the model are optimized for the same pressure
range, using the FFCM-y computation results as the optimization targets. The 12-species skeletal model is called FFCMy-12 mechanism in this report and it is fed into the FlameMaster code for calculation of laminar flames.

\subsection{Numerical Solver}
\label{sec:numsol}

To numerically simulate combustion using the FPV approach, a C++ code is developed based on OpenFOAM 4.0. There is a similar open-source code called flameletFoam \cite{mullerflameletfoam}. However, it only provides the original version of the flamelet model. The progress variable approach is not available. Moreover, tabulated temperature is directly used without any correction, which makes the code applicable only to incompressible combustions. Last but not least, since the code is based on the outdated 2.4 version of OpenFOAM, selection of turbulence model is quite limited. For example, the hybrid RANS/LES based on the SST $k-\omega$ model is not available. In the newly developed code, transport equations \eqref{zequation}, \eqref{varzequation} and \eqref{cequation} are solved for $\widetilde{Z}$, $\widetilde{Z^{''2}}$ and $\widetilde{C}$. These quantities are then accessed to retrieve composition, $\widetilde{\dot{\omega}_{C}}$ and $h_{c}$ from the tables. Since $\widetilde{\dot{\omega}_{C}}$ is tabulated only for background pressure of $200$ bar, it is rescaled using the power law \eqref{omegacvsp} and \eqref{rescalingalpha} for other values of pressure. In this report, the energy equation is solved for $h_{a}$. After that, sensible enthalpy $h_{s}$ is obtained by subtracting $h_{c}$ from $h_{a}$. Temperature is then calculated from sensible enthalpy $h_{s}$ using the Joint Army Navy NASA Air Force (JANAF) polynomials and tables of thermodynamics \cite{nasa1993}. 

In this report, the hybrid RANS/LES based on the SST $k-\omega$ model is selected for both the FPV approach and the one-step-kinetics method. The latter only considers the one-step global chemical reaction and uses the canonical OpenFOAM code called rhoReactingFoam \cite{xiongaiaaj2020}. As temperature for each individual species is higher than its critical point of gas, all species are in the gaseous phase. Hence, the ideal-gas state equation is employed. The reacting-mixtures model is adopted to calculate the properties of the mixture. Since compressibility is considered in the combustion simulation, a density-based thermodynamics package is used. The differencing schemes are second-order accurate in both space and time. Implicit backward differencing is selected for time discretization. For spatial discretization, Gaussian integration is chosen. Second-order derivatives are approximated using linear interpolation from cell centers to cell faces.

\section{Results and Discussions}

\subsection{The Flamelet Solution}
\label{sec:flamelet}

\begin{figure}
    \begin{subfigmatrix}{2}
        \subfigure[S curve]
{\includegraphics{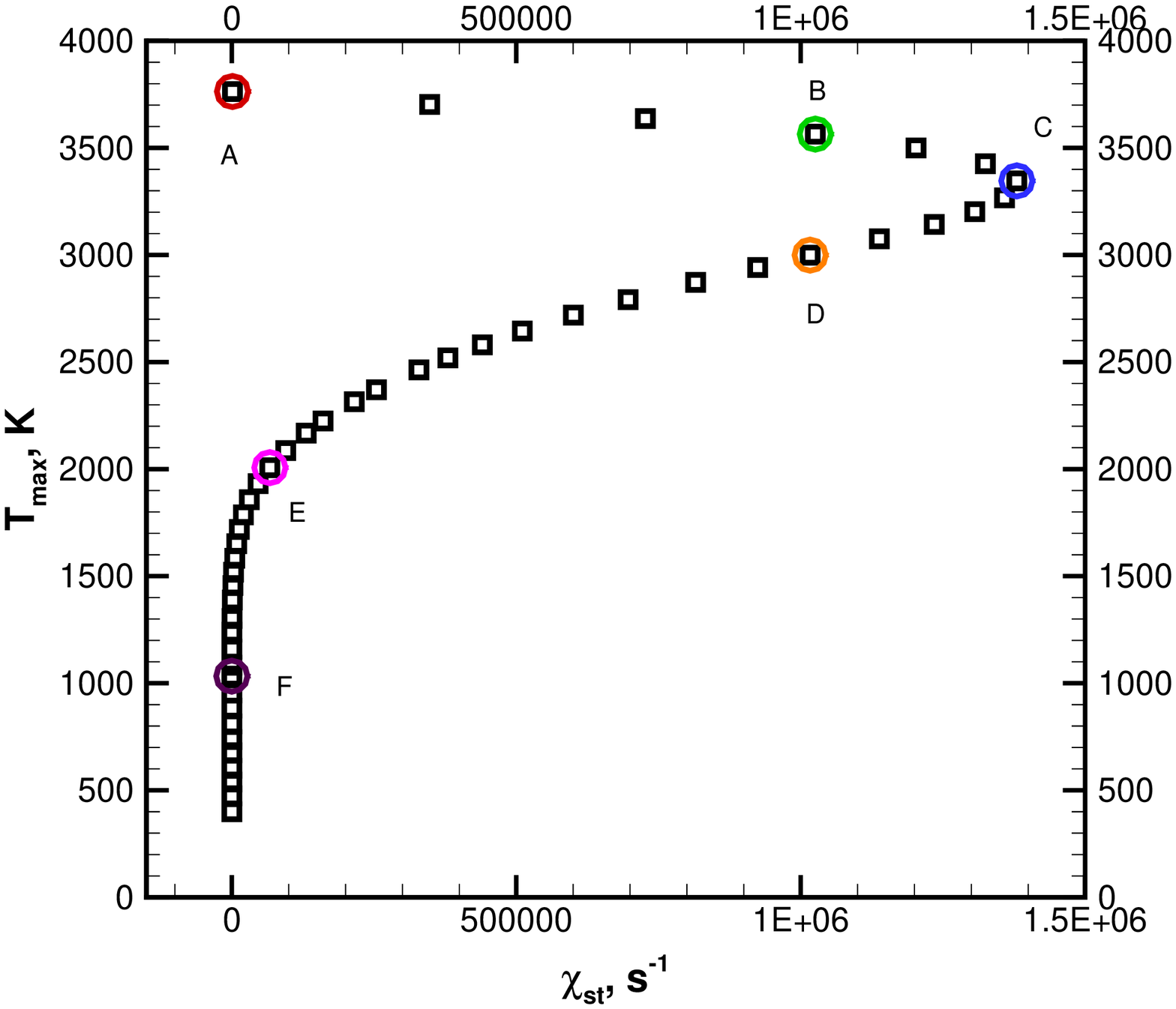}
        \label{fig:scurve}}
        \subfigure[$T_{max}$ vs. $C$]
{\includegraphics{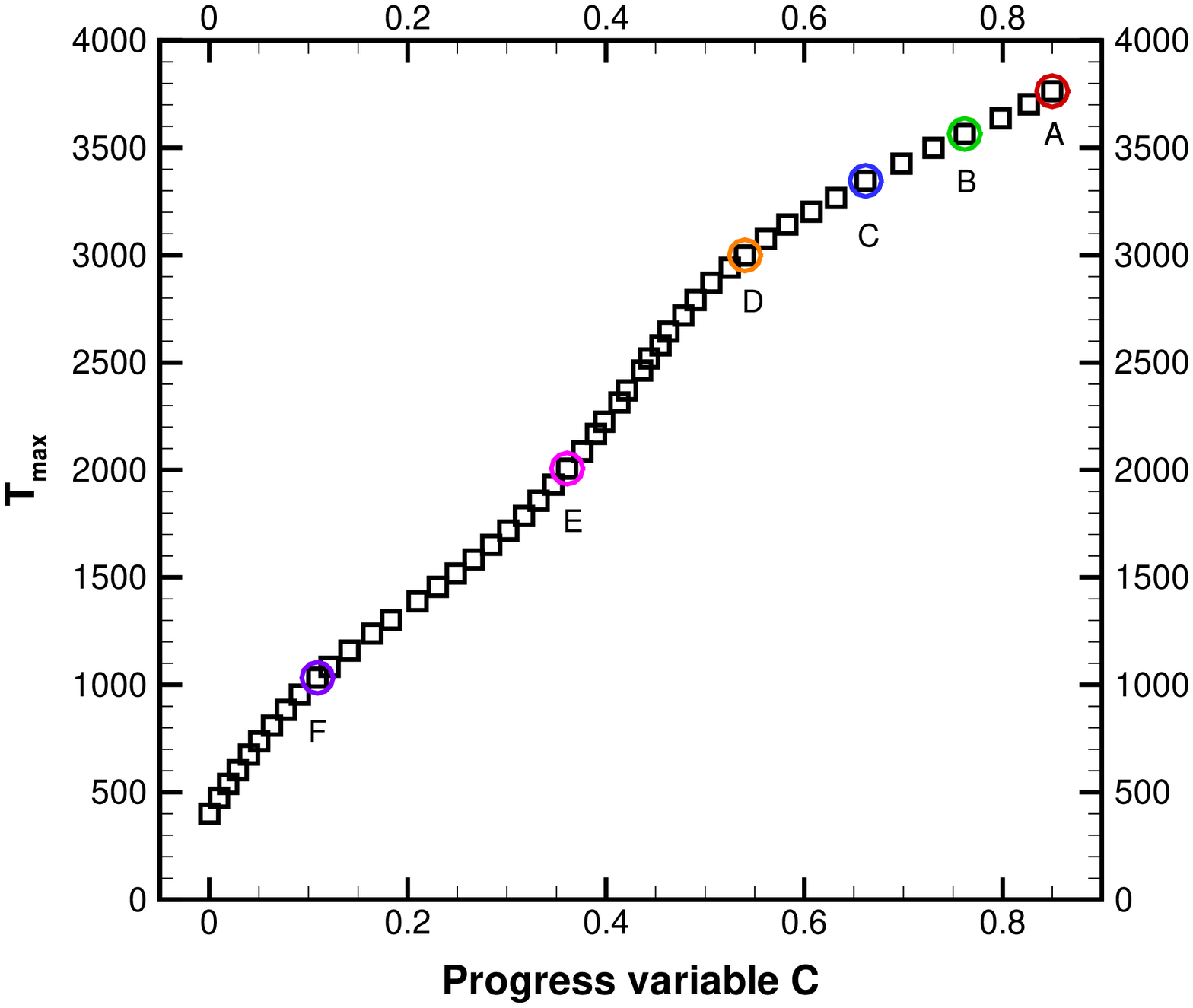}
        \label{fig:tmaxc}}
    \end{subfigmatrix}
    \caption{Solutions of the steady flamelet equations for methane/oxygen combustion with $T_{f} = T_{o} = 400 K$.}
    \label{fig:tmaxflamelet}
\end{figure}

\begin{figure}
    \begin{subfigmatrix}{2}
{\includegraphics{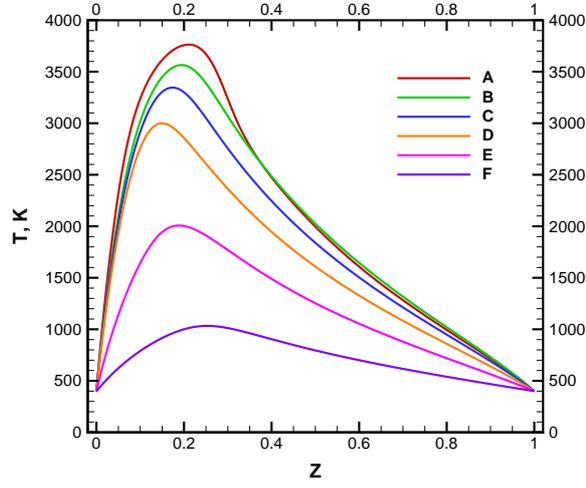}}
    \end{subfigmatrix}
    \caption{Temperature as a function of the mixture fraction (Z).}
    \label{fig:tvszflamelet}
\end{figure}

To apply the FPV model to the turbulent combustion of the ten-injector chamber, steady-state laminar flamelet solutions 
are obtained first. The fuel and oxidizer are pure methane and oxygen, respectively; operating pressure is $200$ bar; and temperature on both fuel and oxidizer sides is $400 K$. Figure~\ref{fig:scurve} shows the maximum temperature in a flamelet solution as a function of $\chi_{st}$, which is frequently called the S curve. It is multi-valued as mentioned above and only the upper branch can be used in the SLF model.
If the progress variable, defined here as the total mass fraction of $CO_{2}$, $H_{2}O$, $CO$ and $H_{2}$, is used as the independent variable, then the curve becomes single-valued and monotonic as shown in Figure~\ref{fig:tmaxc}. Hence, both the upper and lower branches can be explored in CFD.
Selected particular flamelet solutions on the S curve are shown in Figure~\ref{fig:tvszflamelet}.

\begin{figure}
    \begin{subfigmatrix}{2}
        \subfigure[$\omega_{C}$ vs. Z]
{\includegraphics{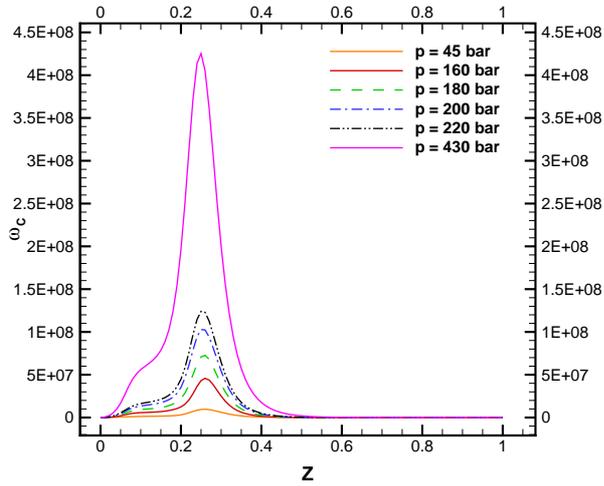}
        \label{fig:omc_Z}}
        \subfigure[$\omega_{C,max}$ vs. p]
{\includegraphics{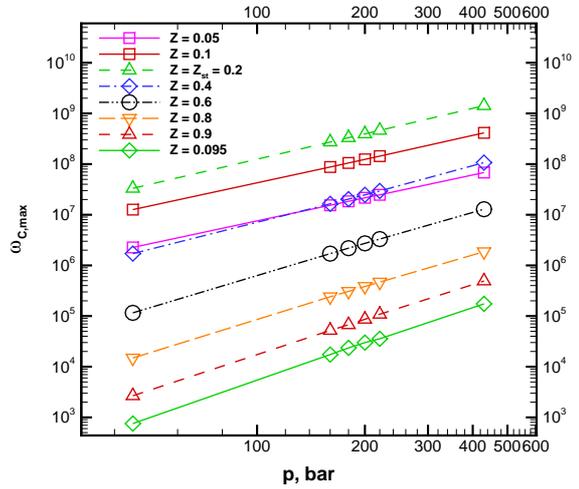}
        \label{fig:maxomc_p}}
    \end{subfigmatrix}
    \caption{Pressure effect on the reaction rate for the progress variable $\omega_{C}$}
    \label{fig:pressureomc}
\end{figure}

In this report, the flamelet solutions are provided for the back pressure of $200$ bar only. Our observation confirms that composition changes little with pressure, whereas the reaction rate for the progress variable ($\omega_{C}$) varies dramatically. Figure~\ref{fig:omc_Z} shows $\omega_{C}$ as a function of $Z$ for different pressures with $C$ fixed at the maximum value $0.85$.

Plotted in a log-log graph, $\omega_{C,max}$ changes linearly with pressure. This indicates $\omega_{C}$ is approximately a power function of pressure, which can be expressed in equation \eqref{omegacvsp} by taking $\omega_{C}$ at $200$ bar as the reference. 

\begin{equation}
\widetilde{\dot{\omega}_{C}} = \widetilde{\dot{\omega}_{C_{0}}}\frac{\tilde{p}^{\alpha}}{\tilde{p_{0}}^{\alpha}}
\label{omegacvsp}
\end{equation}

The power $\alpha$ changes with $Z$ and its variance $Z^{''}$. For a given combination of $Z$ and $Z^{''}$, $\alpha$ can be determined by power regression analysis and by assuming $\alpha$ as a quadratic function of $Z$ and a linear function of $Z^{''}$, $\omega_{C}$ at pressure other than $200$ bar is calculated with the best possible accuracy. The final approximation of $\alpha$ is given in equation \eqref{rescalingalpha}.

\begin{equation}
\alpha = (3.0882Z^{''}+0.0101)Z^{2}+(-3.7851Z^{''}+0.9567)Z+(0.6638Z^{''}+1.4589)
\label{rescalingalpha}
\end{equation}

\subsection{Combustion instability in a multi-injector rocket engine}
\label{sec:combusinsta}

\begin{figure}
    \begin{subfigmatrix}{2}
        \subfigure[Isometric view]
{\includegraphics{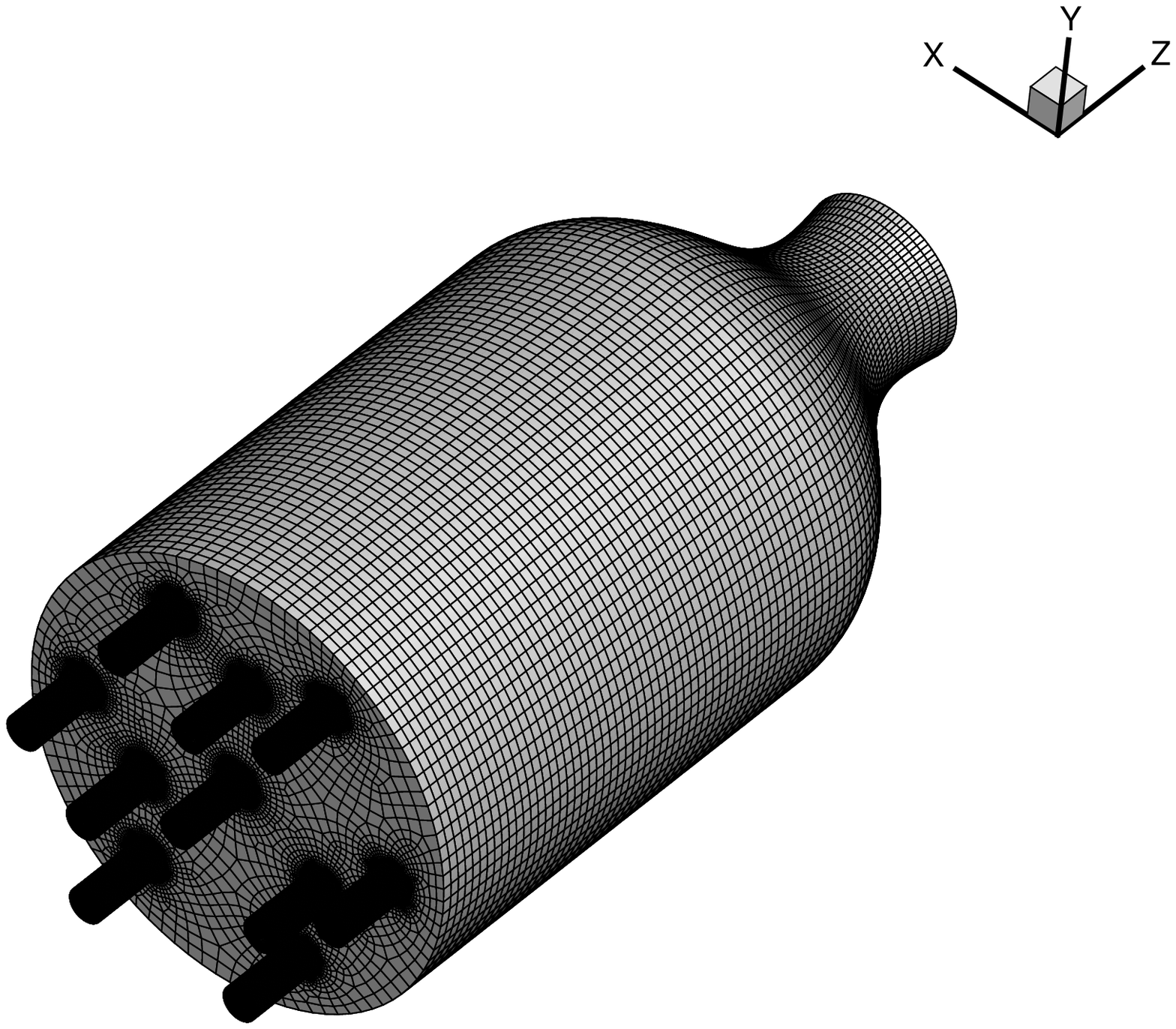}
        \label{fig:3dmesh}}
        \subfigure[Front view]
{\includegraphics{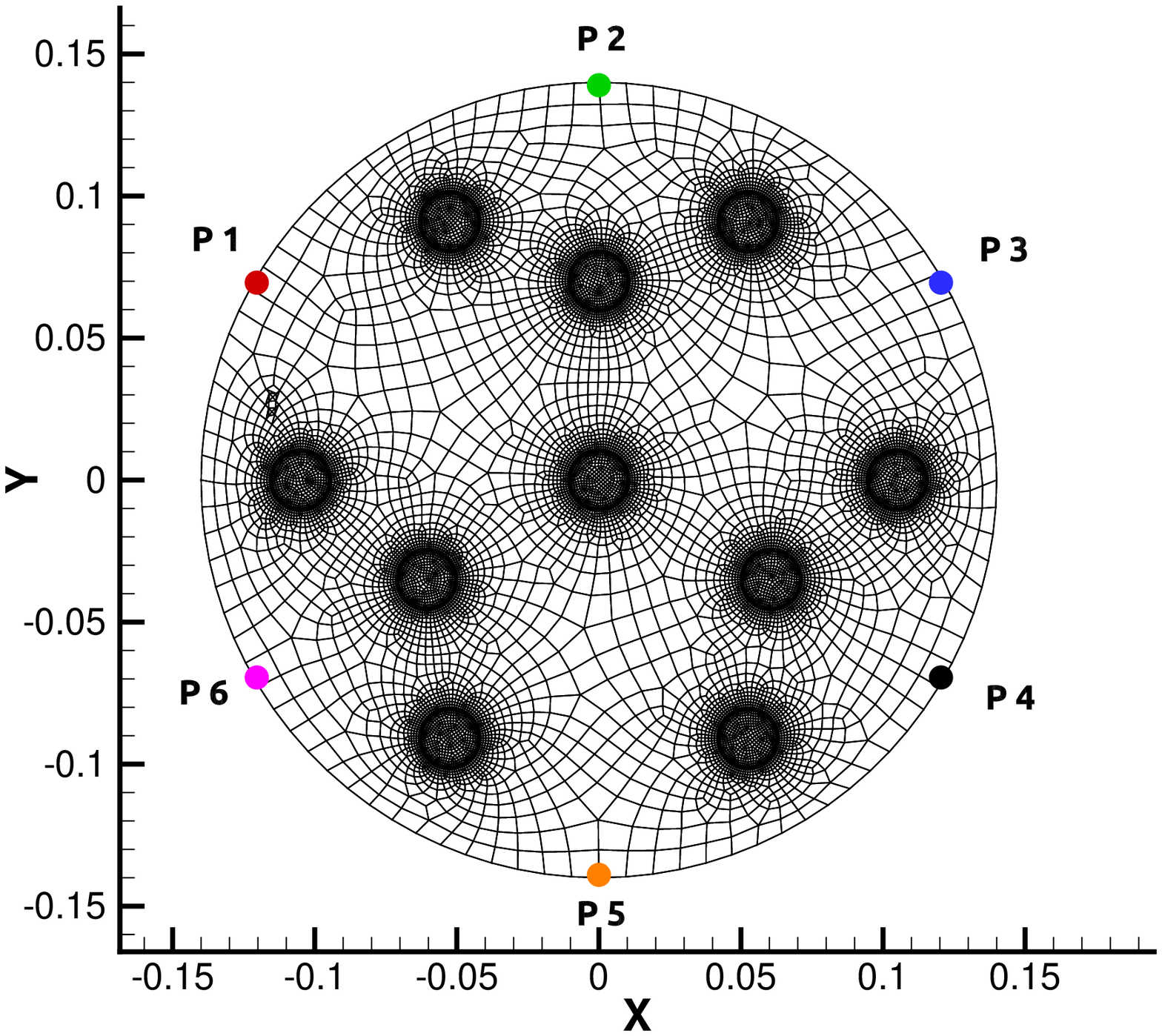}
        \label{fig:2dmesh}}
    \end{subfigmatrix}
    \caption{Unstructured mesh for the ten-injector combustion chamber}
    \label{fig:mesh}
\end{figure}

In this report, turbulent combustion is numerically simulated in a chamber with an convergent-divergent nozzle and ten coaxial fuel/oxidizer injectors\cite{xiongaiaaj2020}. The three-dimension unstructured mesh with $1.608502 \times 10^{6}$ grid cells is shown in Figure~\ref{fig:3dmesh}. The combustion chamber has a diameter of $28$ cm and a length of $53$ cm. The distance between the injector plate and the choked nozzle is $33$ cm. The oxidizer, which is pure oxygen, is injected in the center of each coaxial port. The fuel, which is pure methane, is injected on the outside annulus. The mass fluxes for the fuel and oxidizer are $15$ kg/s and $60$ kg/s, respectively. The diameter of the fuel injector is $1.96$ cm and that of the oxidizer is $2.2$ cm. For the given mass flux, this port geometry makes the mean injection velocity ratio between the fuel and the oxidizer $2.28$. Temperature at both the fuel and oxidizer injector inlets is $400$ K. The initial pressure in the combustion chamber is $200$ bar. 

\begin{figure}
    \begin{subfigmatrix}{2}
        \subfigure[FPV]
{\includegraphics{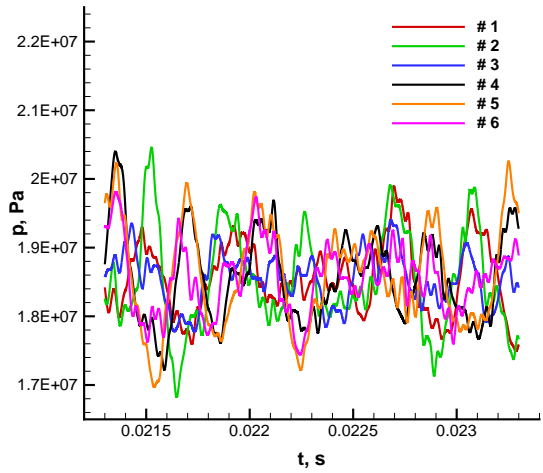}
        \label{fig:p_original_0dot01m_fpv}}
        \subfigure[one step kinetics]
{\includegraphics{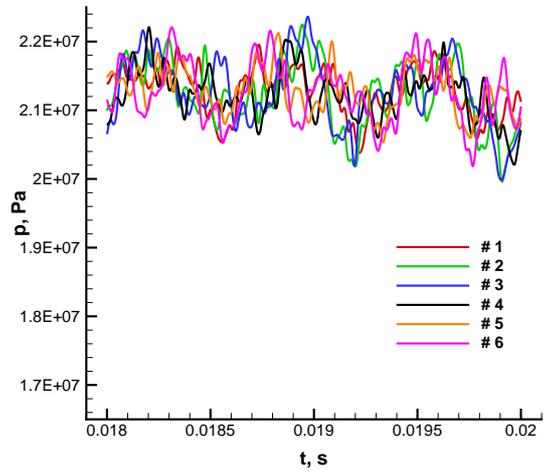}
        \label{fig:p_original_0dot01m_osk}}
    \end{subfigmatrix}
    \caption{Pressure time histories on near-wall probles at $z = 0.01$ m}
    \label{fig:p_original_0dot01m}
\end{figure}

\begin{figure}
    \begin{subfigmatrix}{2}
        \subfigure[FPV]
{\includegraphics{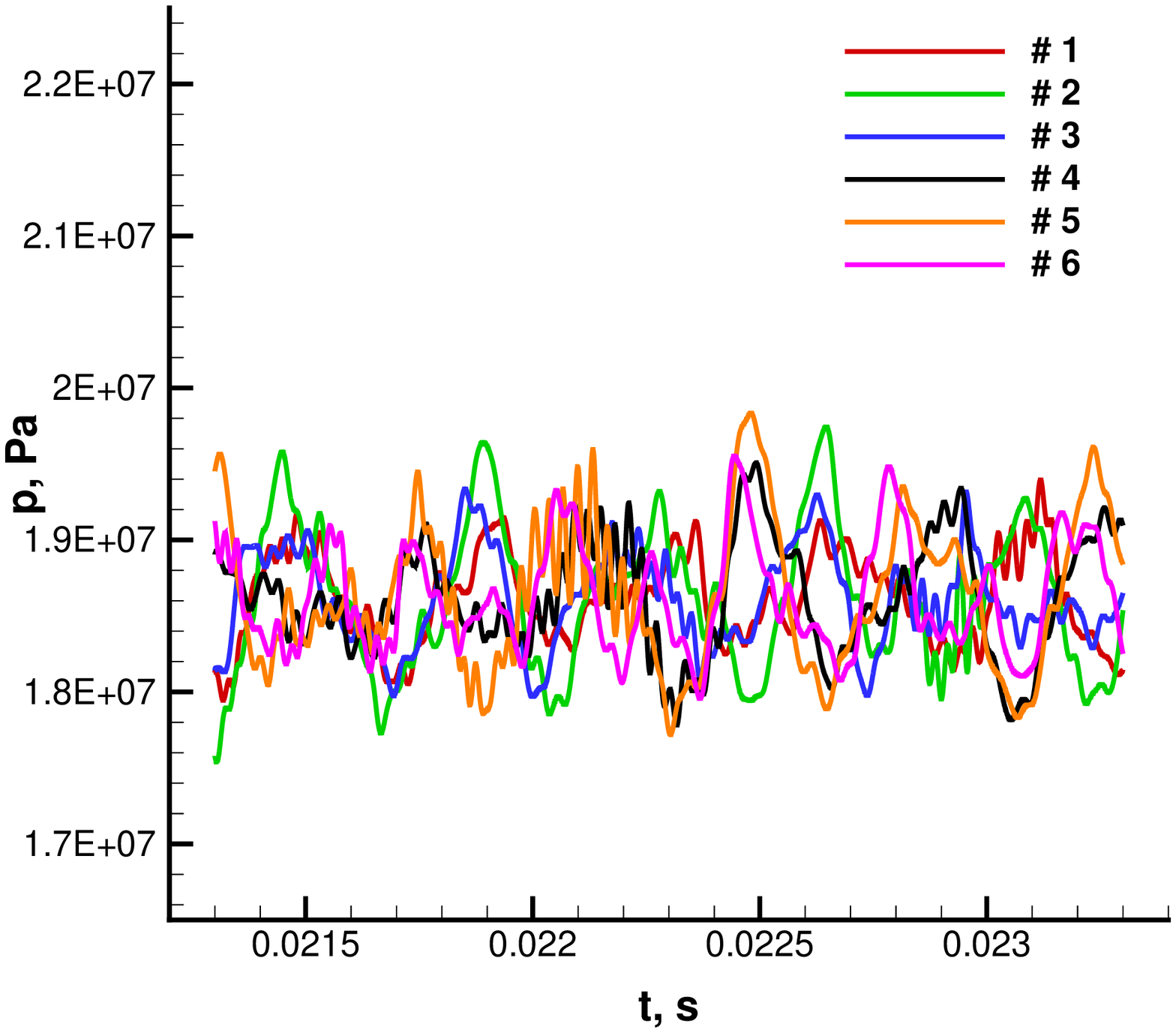}
        \label{fig:p_original_0dot18m_fpv}}
        \subfigure[one step kinetics]
{\includegraphics{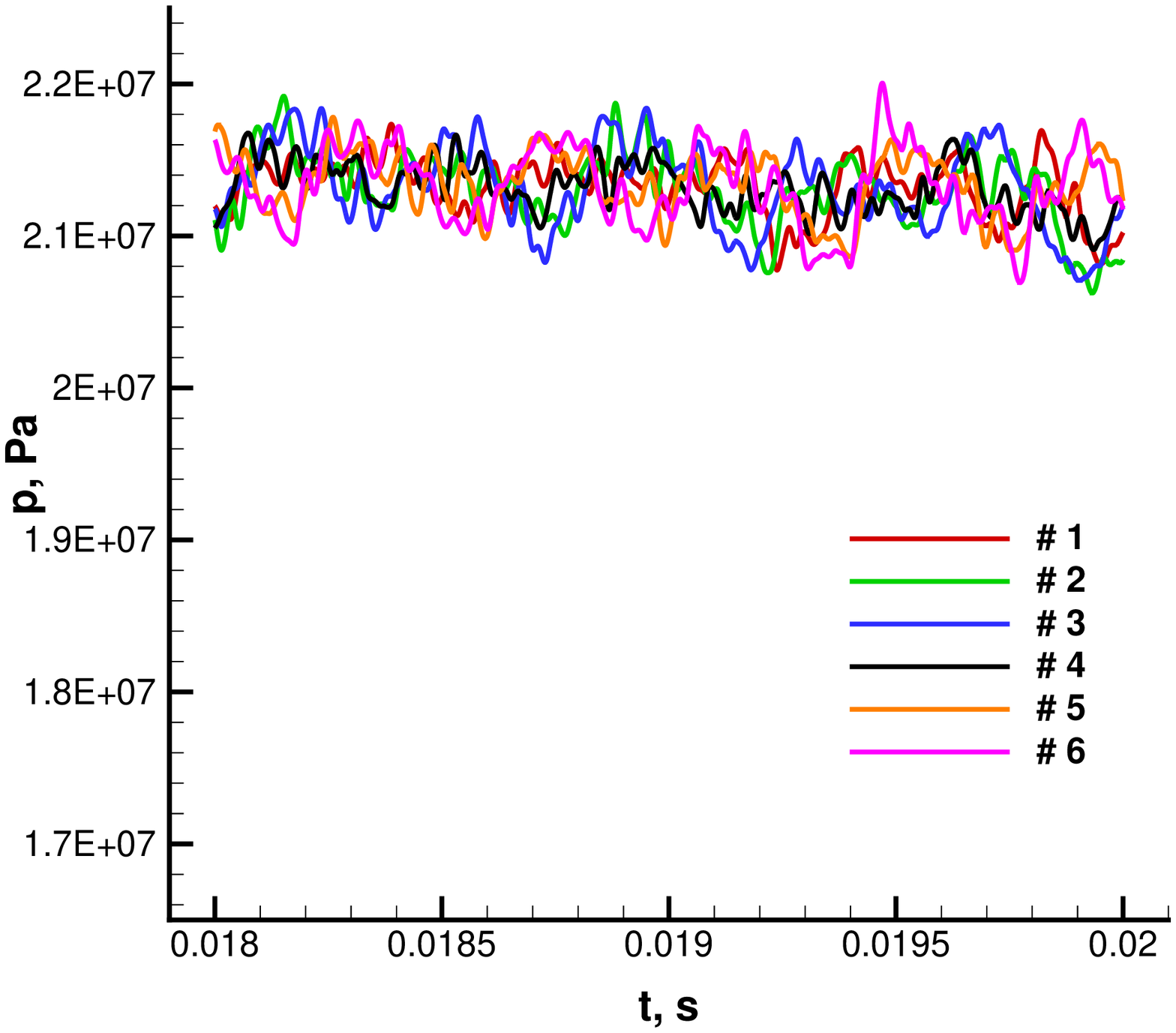}
        \label{fig:p_original_0dot18m_osk}}
    \end{subfigmatrix}
    \caption{Pressure time histories on near-wall probles at $z = 0.18$ m}
    \label{fig:p_original_0dot18m}
\end{figure}

\begin{figure}
    \begin{subfigmatrix}{2}
        \subfigure[FPV]
{\includegraphics{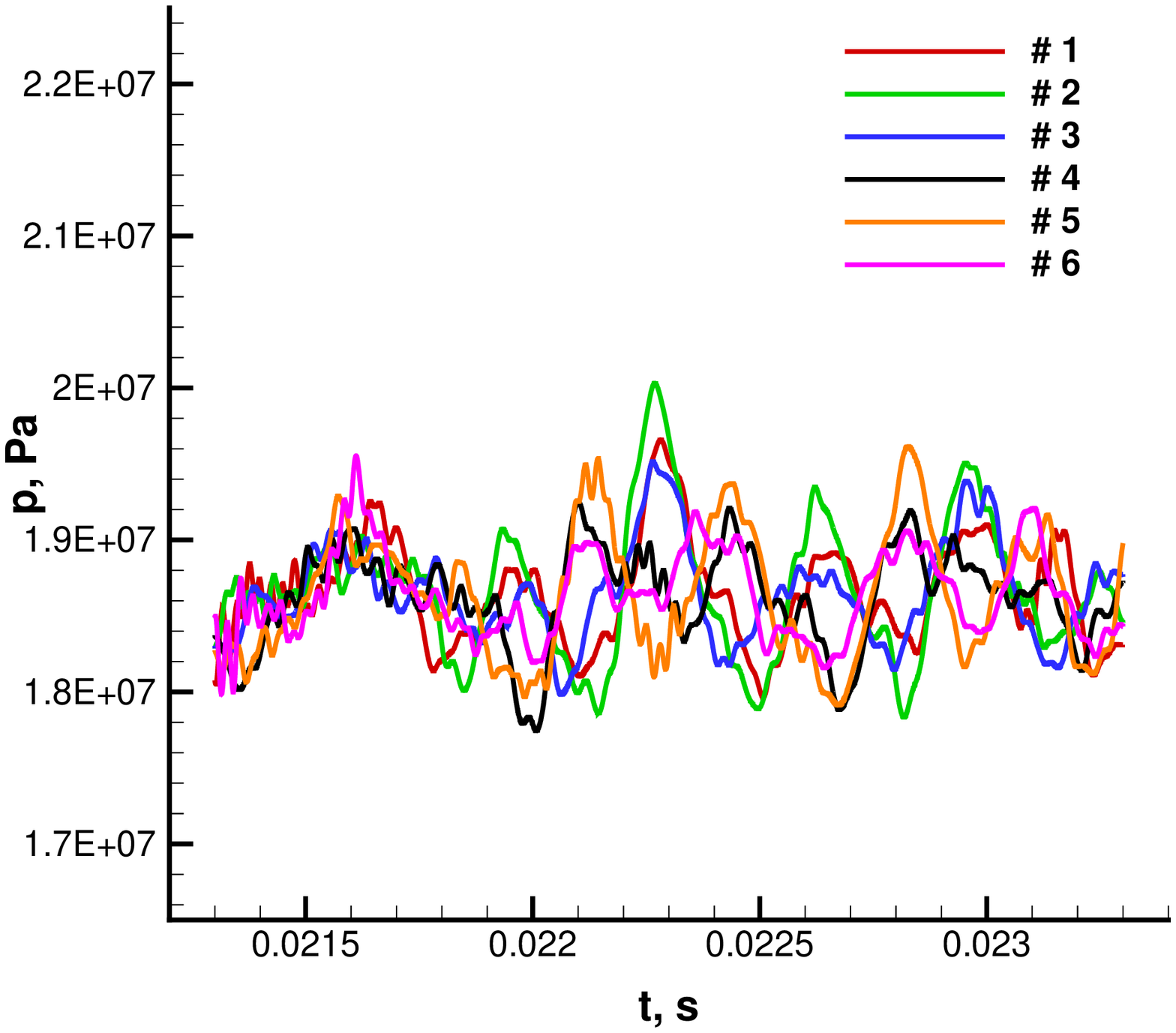}
        \label{fig:p_original_0dot33m_fpv}}
        \subfigure[one step kinetics]
{\includegraphics{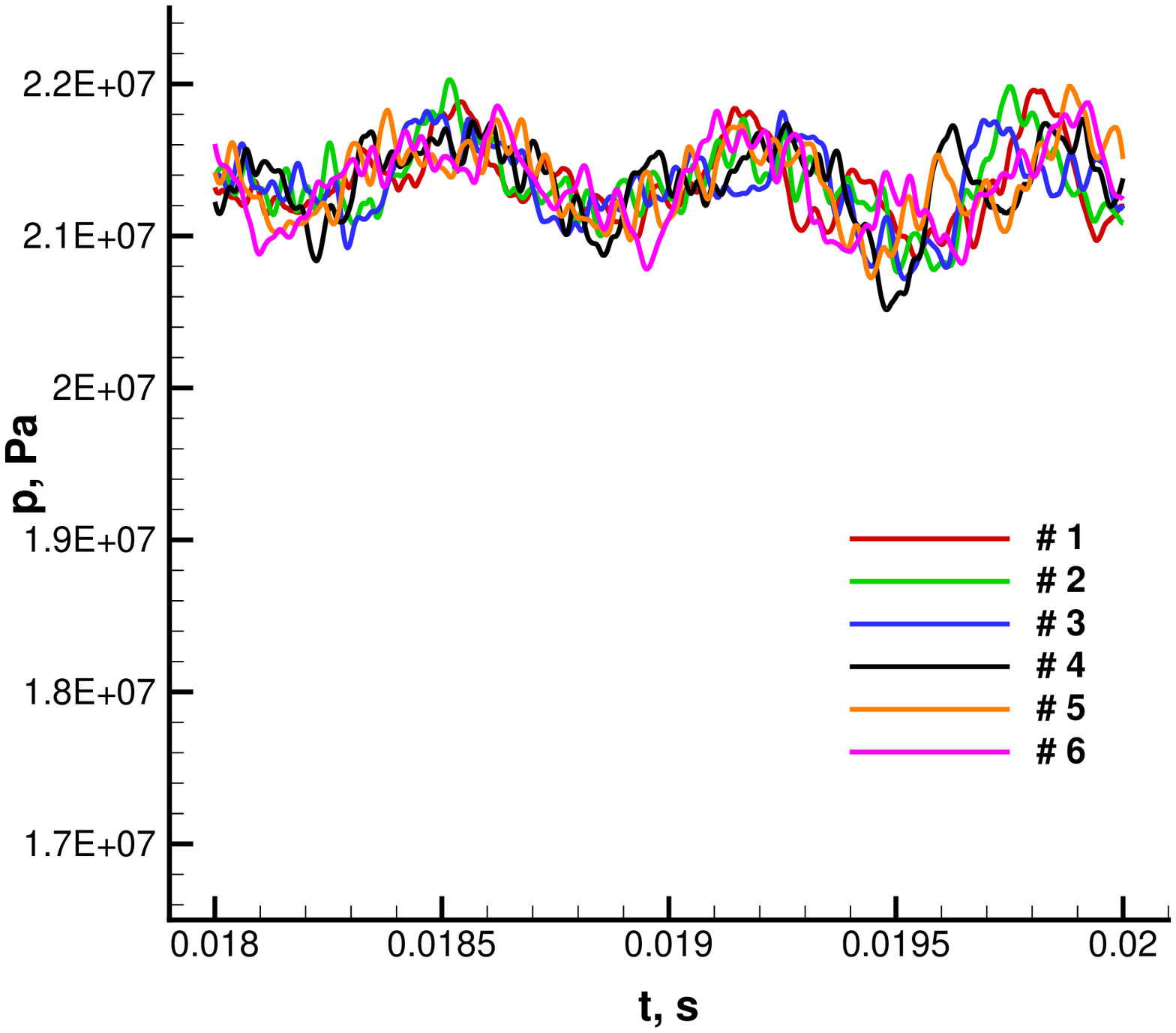}
        \label{fig:p_original_0dot33m_osk}}
    \end{subfigmatrix}
    \caption{Pressure time histories on near-wall probles at $z = 0.33$ m}
    \label{fig:p_original_0dot33m}
\end{figure}

\begin{figure}
    \begin{subfigmatrix}{2}
        \subfigure[FPV]
{\includegraphics{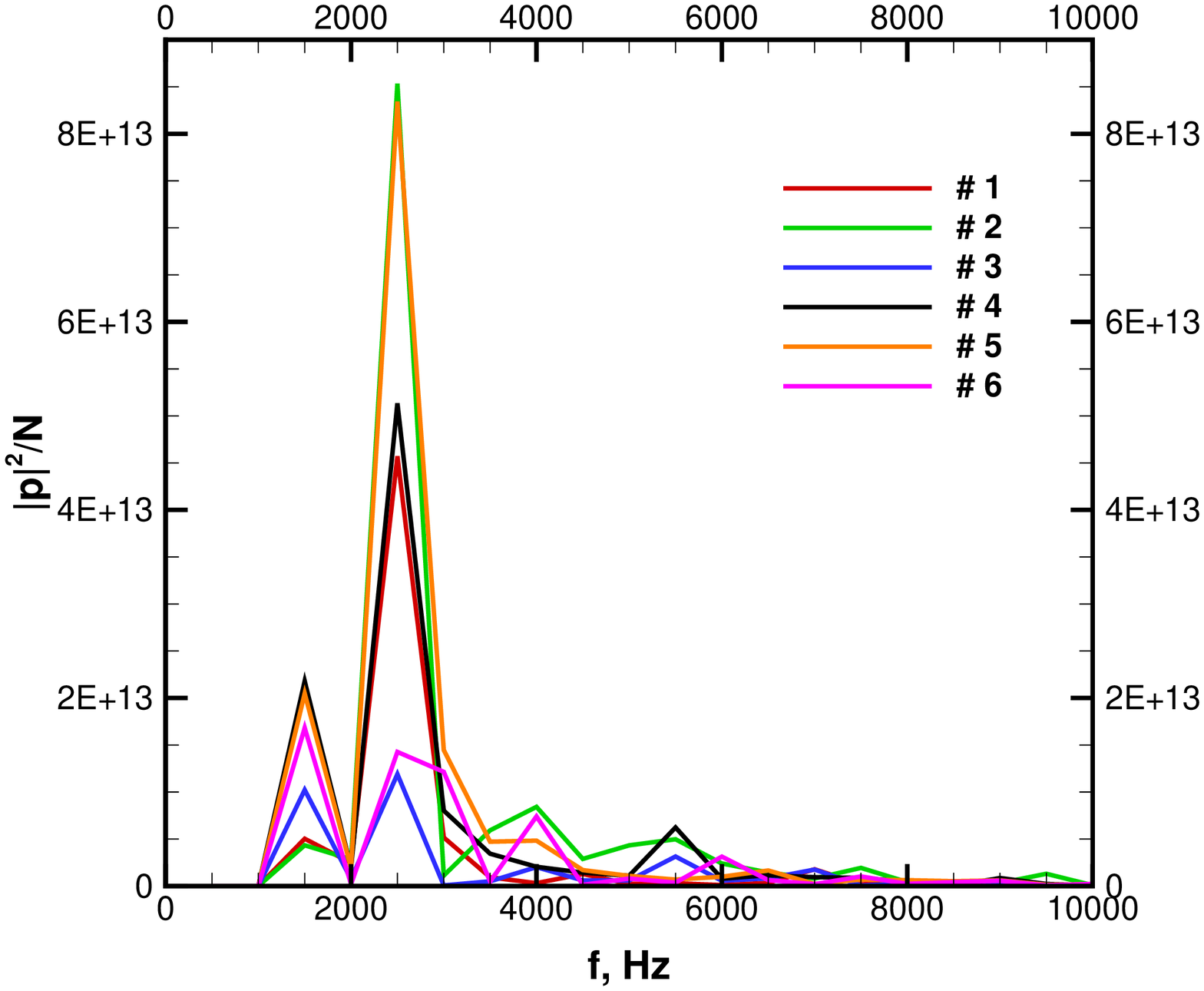}
        \label{fig:p_spectrum_0dot01m_fpv}}
        \subfigure[one step kinetics]
{\includegraphics{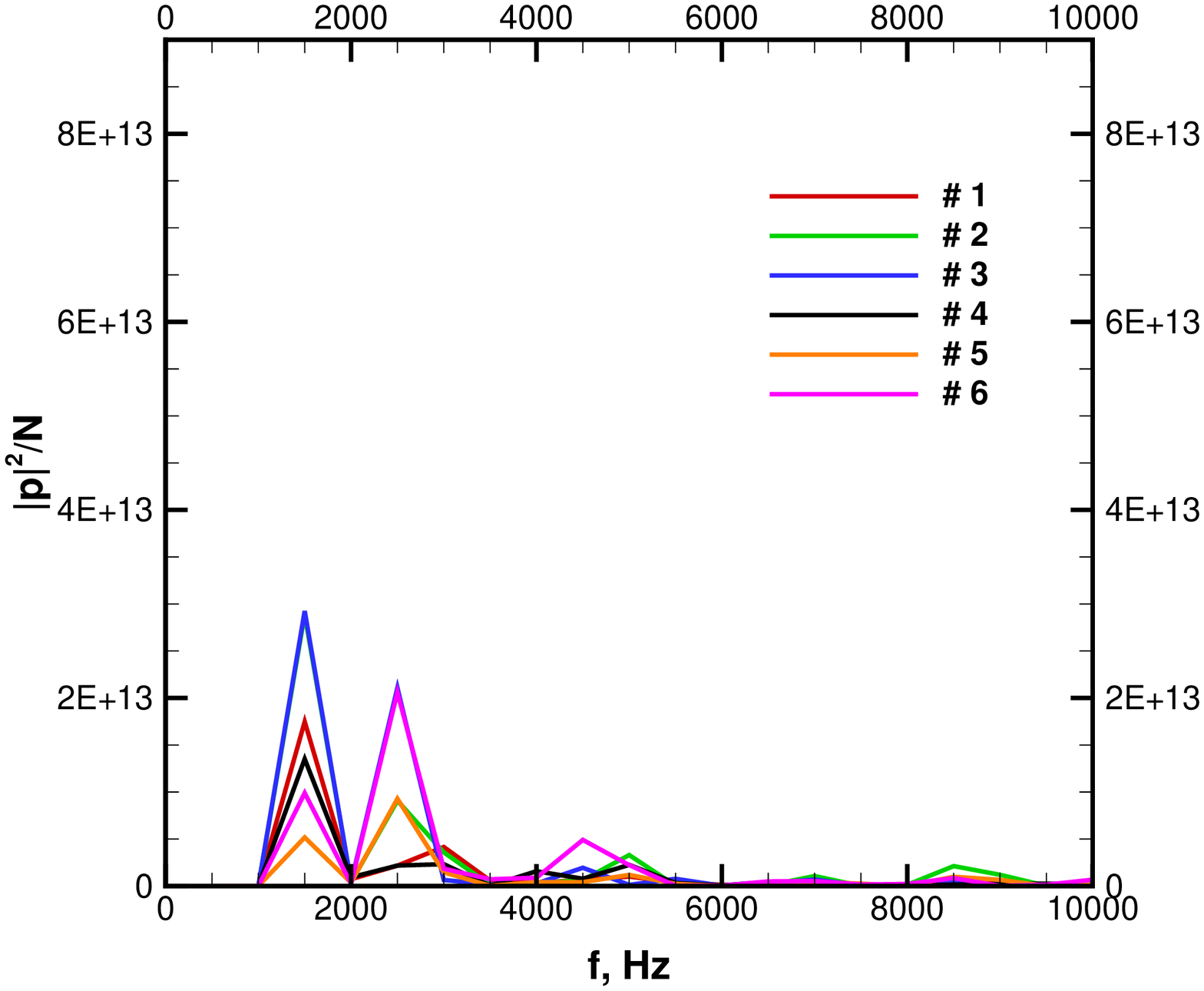}
        \label{fig:p_spectrum_0dot01m_osk}}
    \end{subfigmatrix}
    \caption{Pressure spectrum on near-wall probles at $z = 0.01$ m}
    \label{fig:p_spectrum_0dot01m}
\end{figure}

\begin{figure}
    \begin{subfigmatrix}{2}
        \subfigure[FPV]
{\includegraphics{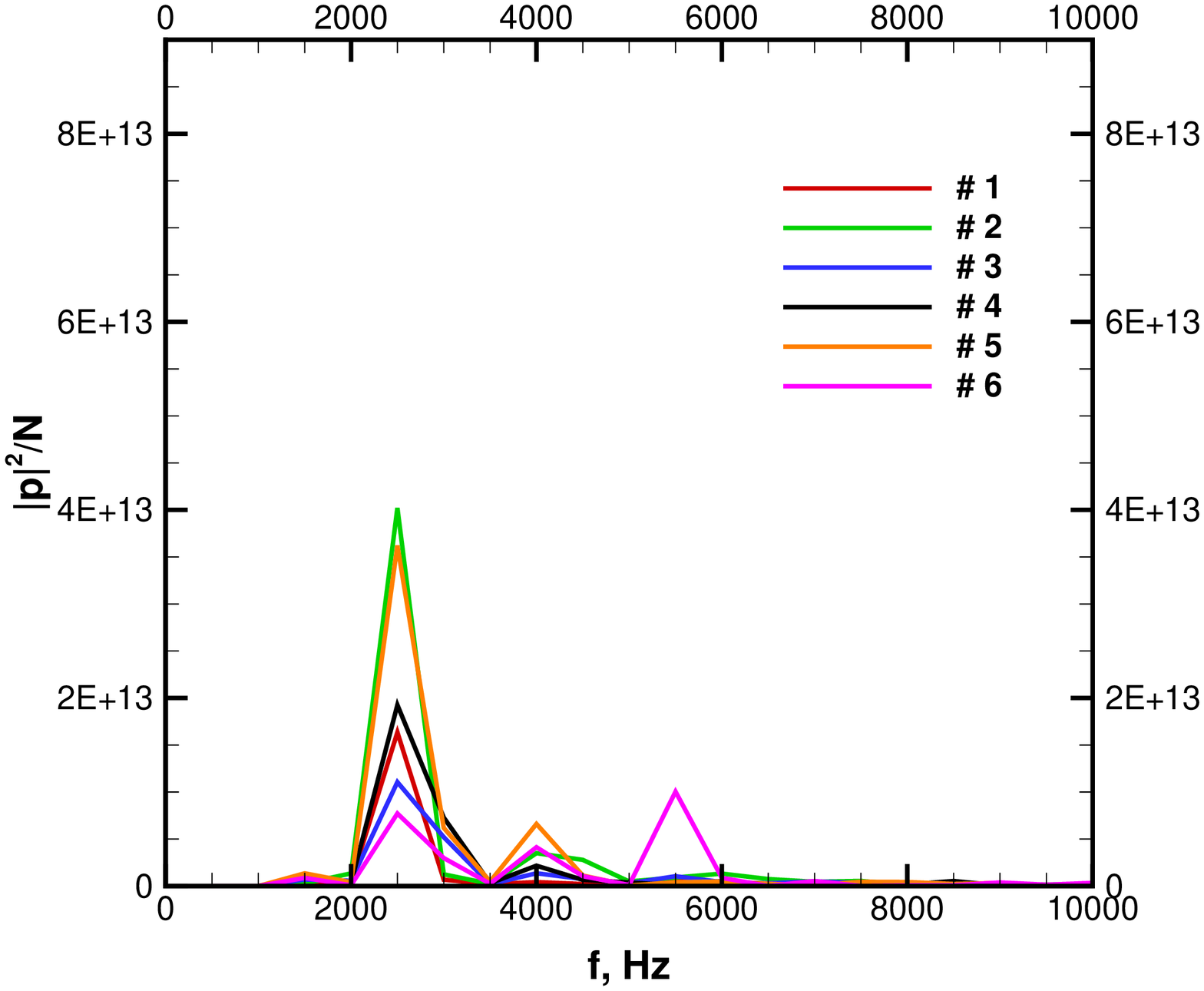}
        \label{fig:p_spectrum_0dot18m_fpv}}
        \subfigure[one step kinetics]
{\includegraphics{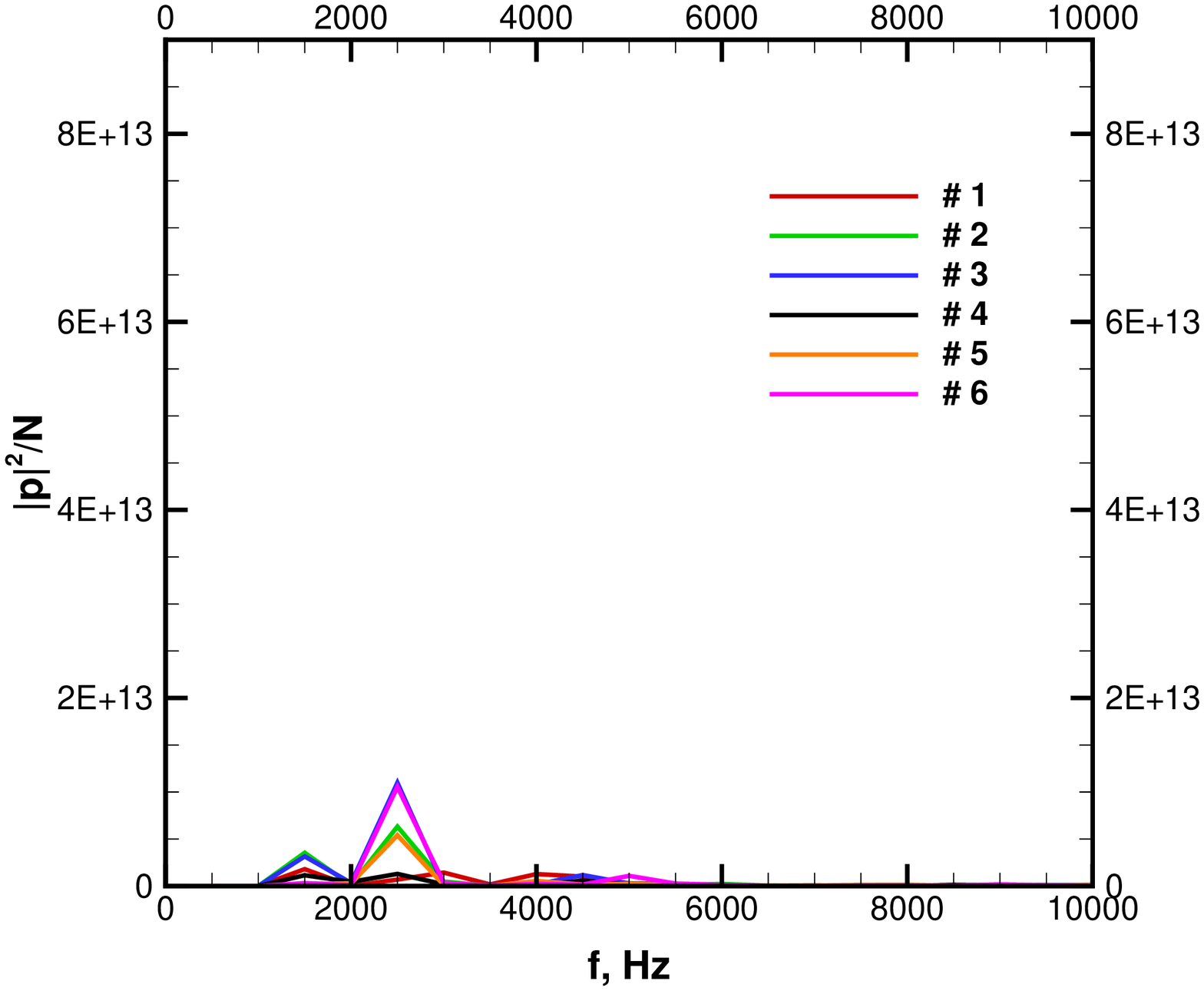}
        \label{fig:p_spectrum_0dot18m_osk}}
    \end{subfigmatrix}
    \caption{Pressure spectrum on near-wall probles at $z = 0.18$ m}
    \label{fig:p_spectrum_0dot18m}
\end{figure}

\begin{figure}
    \begin{subfigmatrix}{2}
        \subfigure[FPV]
{\includegraphics{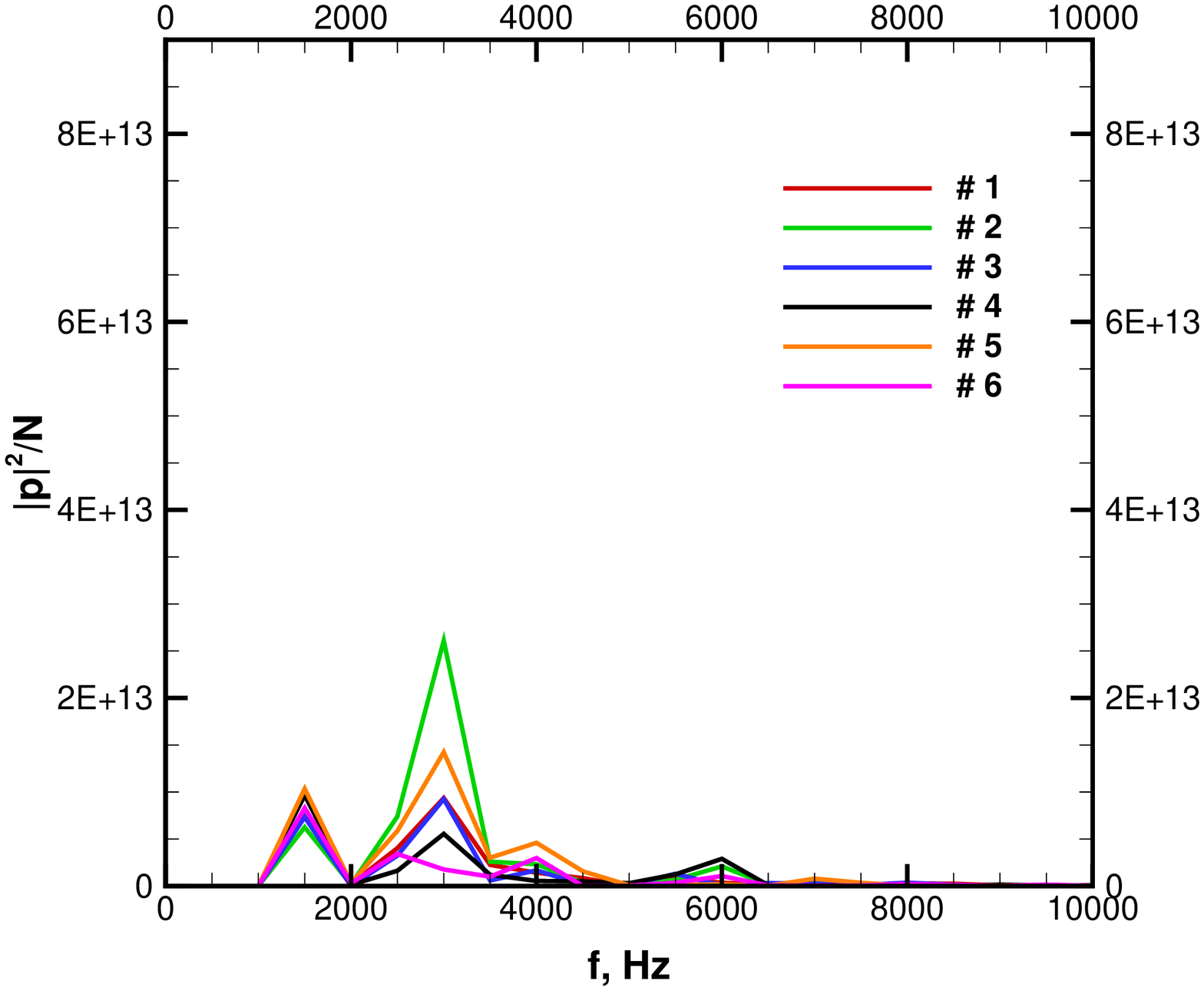}
        \label{fig:p_spectrum_0dot33m_fpv}}
        \subfigure[one step kinetics]
{\includegraphics{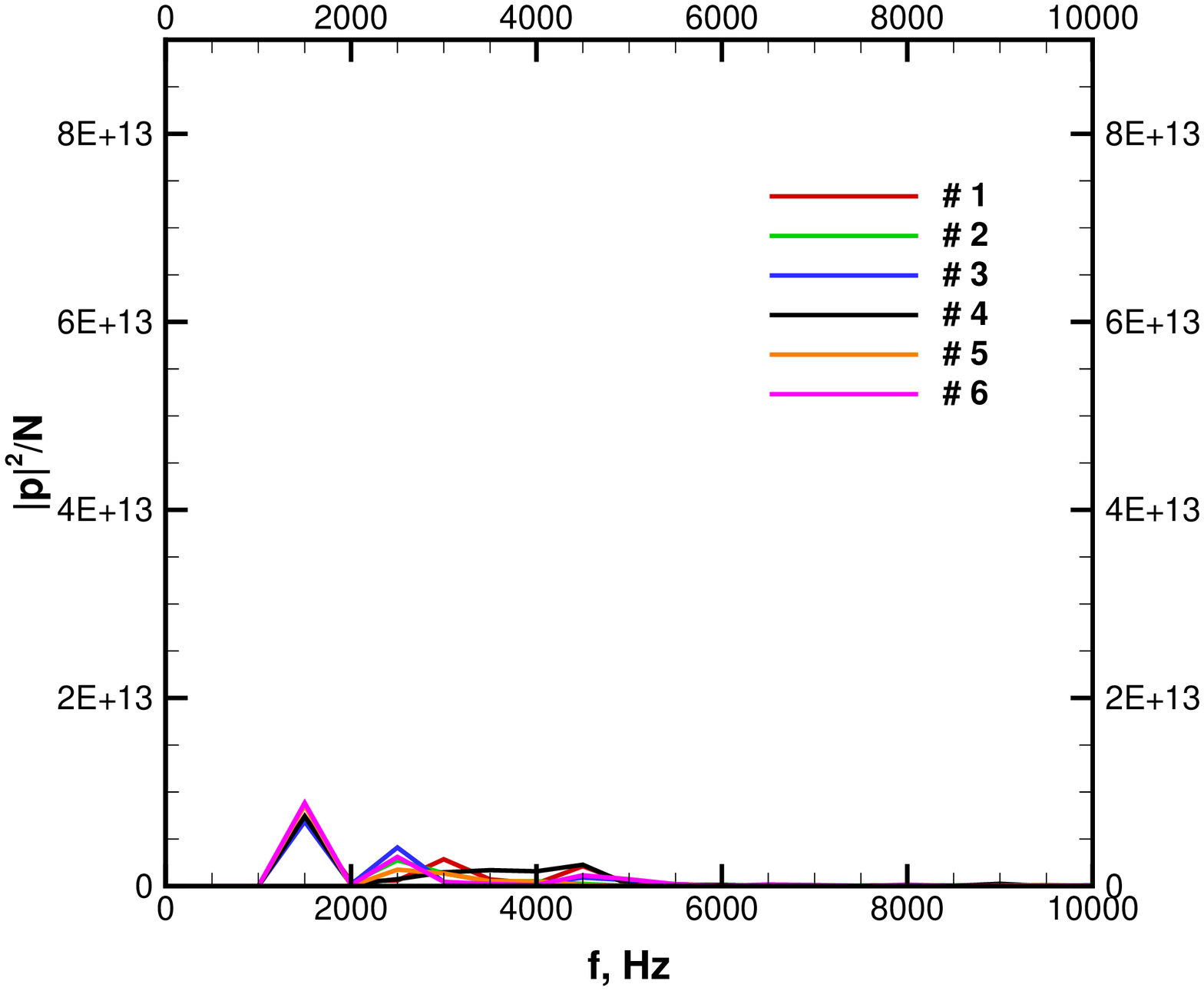}
        \label{fig:p_spectrum_0dot33m_osk}}
    \end{subfigmatrix}
    \caption{Pressure spectrum on near-wall probles at $z = 0.33$ m}
    \label{fig:p_spectrum_0dot33m}
\end{figure}

Combustion in the chamber described above is simulated separately using the newly-developed OpenFOAM-based code for the FPV approach and the canonical rhoReactingFoam code for the one step kinetics (OSK) method. To monitor time histories of pressure and other quantities for study of combustion instability, probes are placed $1$ mm away from the chamber wall and equally distributed in the circumferential direction as shown in Figure~\ref{fig:2dmesh}. Such probes are set up at three different locations downstream the injector plate, which are $z = 0.01$ m, $z = 0.18$ m and $z = 0.33$ m. Figures~\ref{fig:p_original_0dot01m}, \ref{fig:p_original_0dot18m} and \ref{fig:p_original_0dot33m} show the pressure time histories on the near-wall probes. The FPV approach predicts a chamber pressure nearly $30$ bar lower than the OSK method. By theoretical estimation, the flame temperature in equilibrium of the 12-species mechanism (used in the FPV approach) is around $3800$ K while that of the 4-species one-step global reaction rises to $5000$ K. This indicates the flames simulated by the FPV approach is cooler in the combustion chamber, which is consistent with a calculated lower pressure. Figures~\ref{fig:p_spectrum_0dot01m}, \ref{fig:p_spectrum_0dot18m} and \ref{fig:p_spectrum_0dot33m} show the pressure spectrum on the near-wall probes. The time window for the Fourier analysis is $[0.0213s, 0.0233s]$ in the FPV approach and $[0.018s, 0.02s]$ in the OSK method. The same dominant modes are identified for both computations. The corresponding frequencies are $1500$ Hz and $2500$ Hz. Moreover, the amplitude of the $1500$ Hz mode in the FPV approach is comparable to that in the OSK method. However, the mode of $2500$ Hz is stronger in the FPV approach. Following analysis shows that the mode of $1500$ Hz is a longitudinal mode, while the mode of $2500$ Hz is a tangential standing wave.

\begin{figure}
    \begin{subfigmatrix}{2}
        \subfigure[FPV]
{\includegraphics{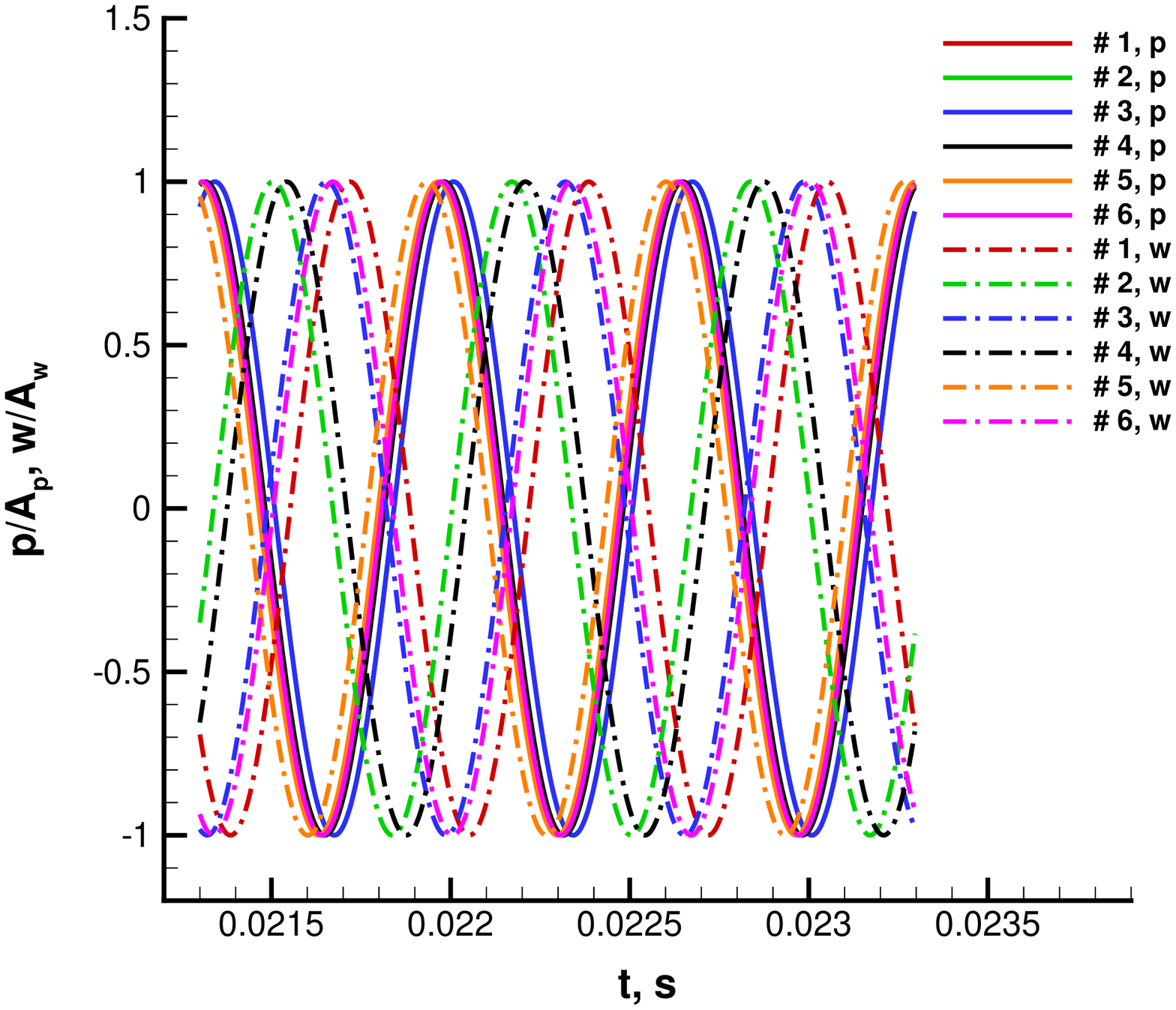}
        \label{fig:p_w_1500hz_0dot01m_fpv}}
        \subfigure[one step kinetics]
{\includegraphics{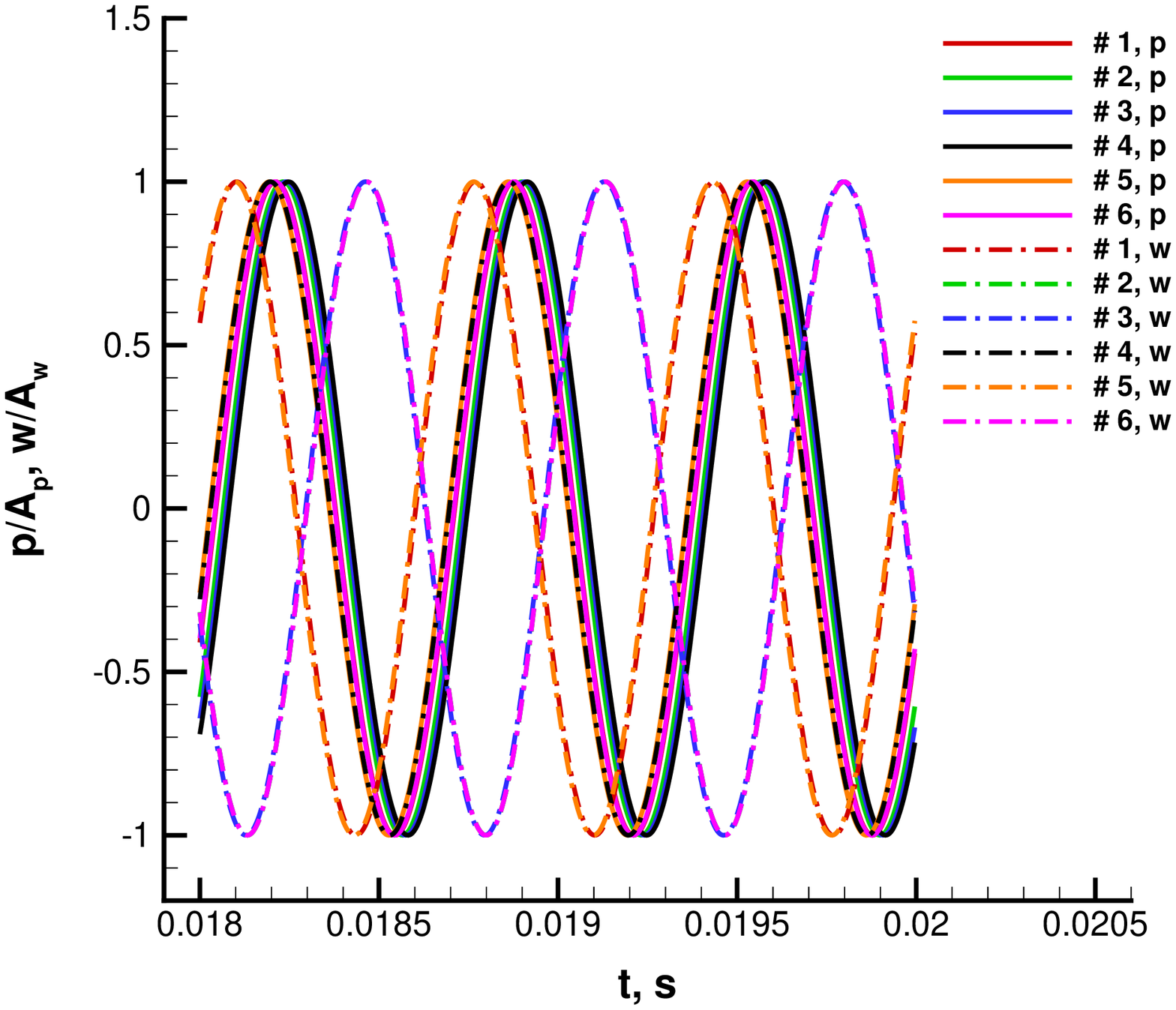}
        \label{fig:p_w_1500hz_0dot01m_osk}}
    \end{subfigmatrix}
    \caption{Rescaled time histories of pressure and axial velocity for the mode of $1500$ Hz at $z = 0.01$ m}
    \label{fig:p_w_1500hz_0dot01m}
\end{figure}

\begin{figure}
    \begin{subfigmatrix}{2}
        \subfigure[FPV]
{\includegraphics{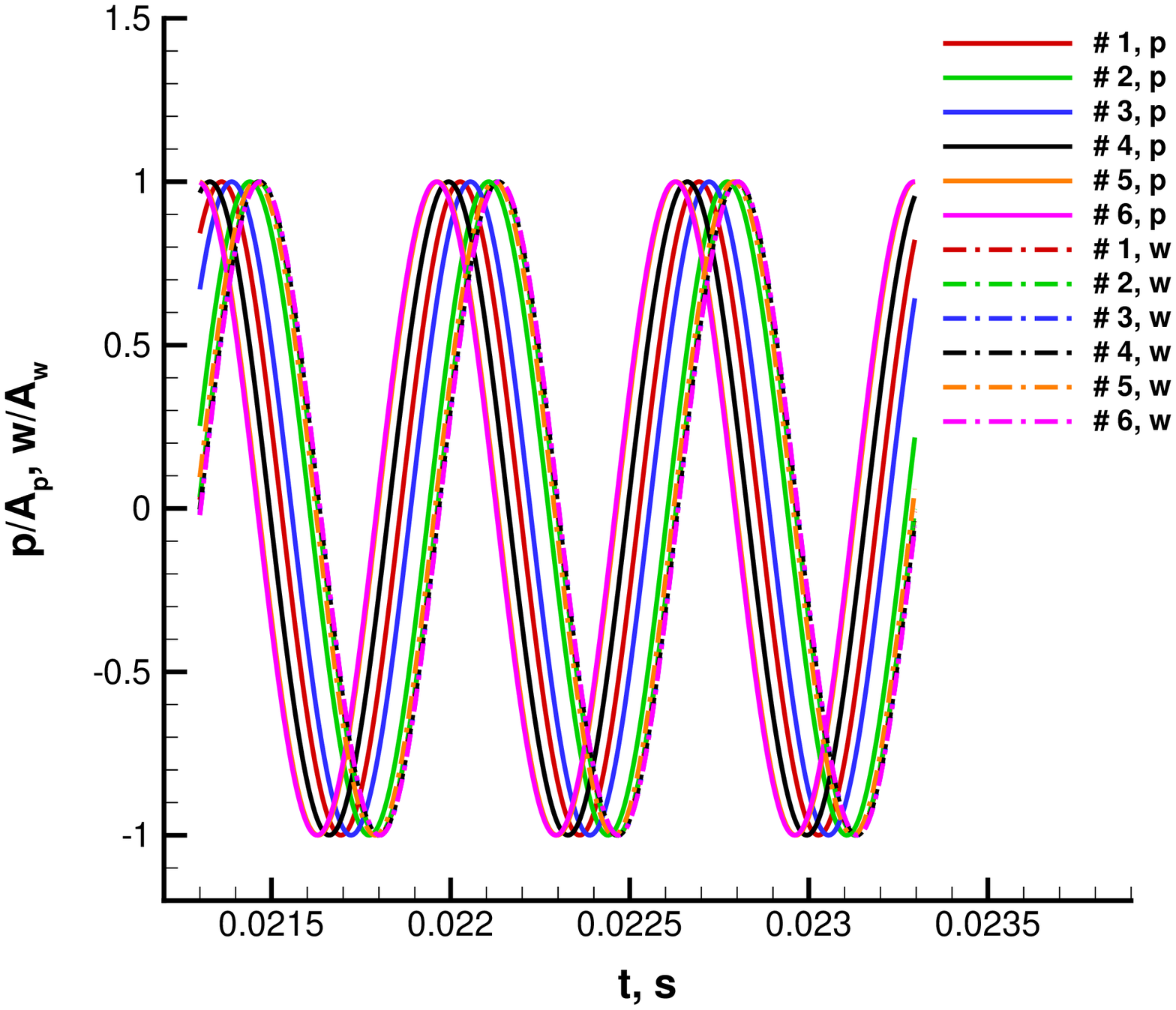}
        \label{fig:p_w_1500hz_0dot18m_fpv}}
        \subfigure[one step kinetics]
{\includegraphics{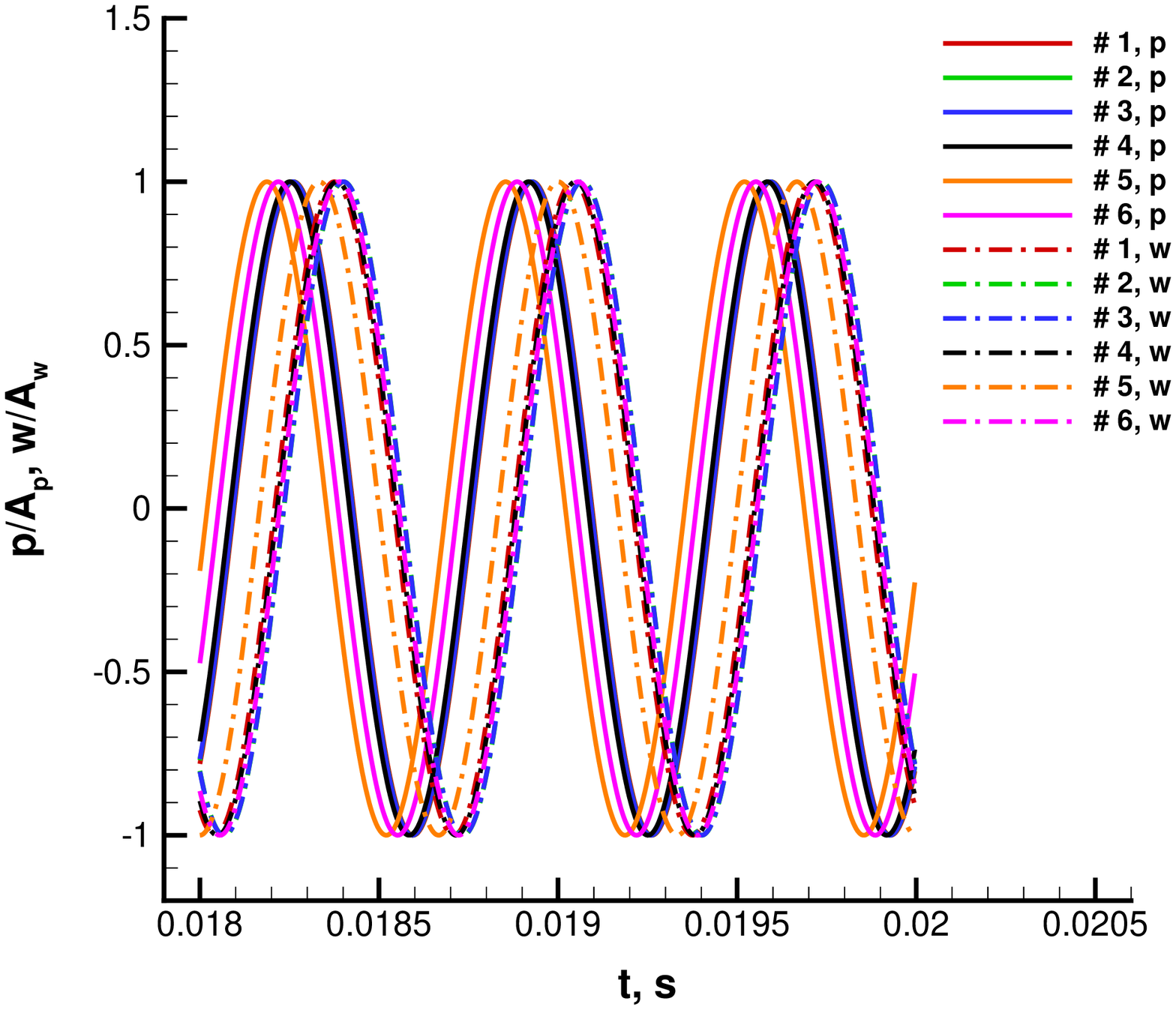}
        \label{fig:p_w_1500hz_0dot18m_osk}}
    \end{subfigmatrix}
    \caption{Rescaled time histories of pressure and axial velocity for the mode of $1500$ Hz at $z = 0.18$ m}
    \label{fig:p_w_1500hz_0dot18m}
\end{figure}

\begin{figure}
    \begin{subfigmatrix}{2}
        \subfigure[FPV]
{\includegraphics{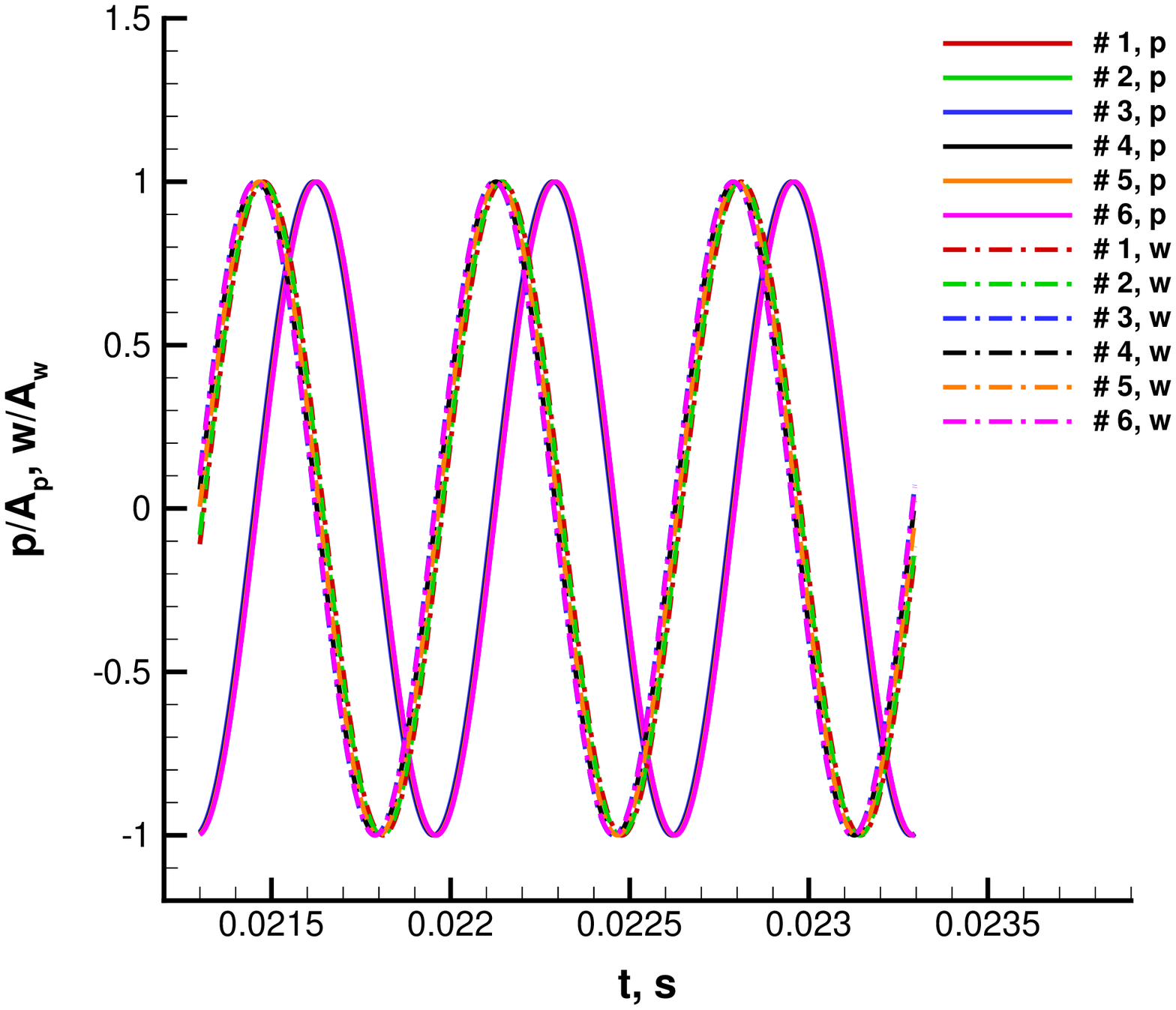}
        \label{fig:p_w_1500hz_0dot33m_fpv}}
        \subfigure[one step kinetics]
{\includegraphics{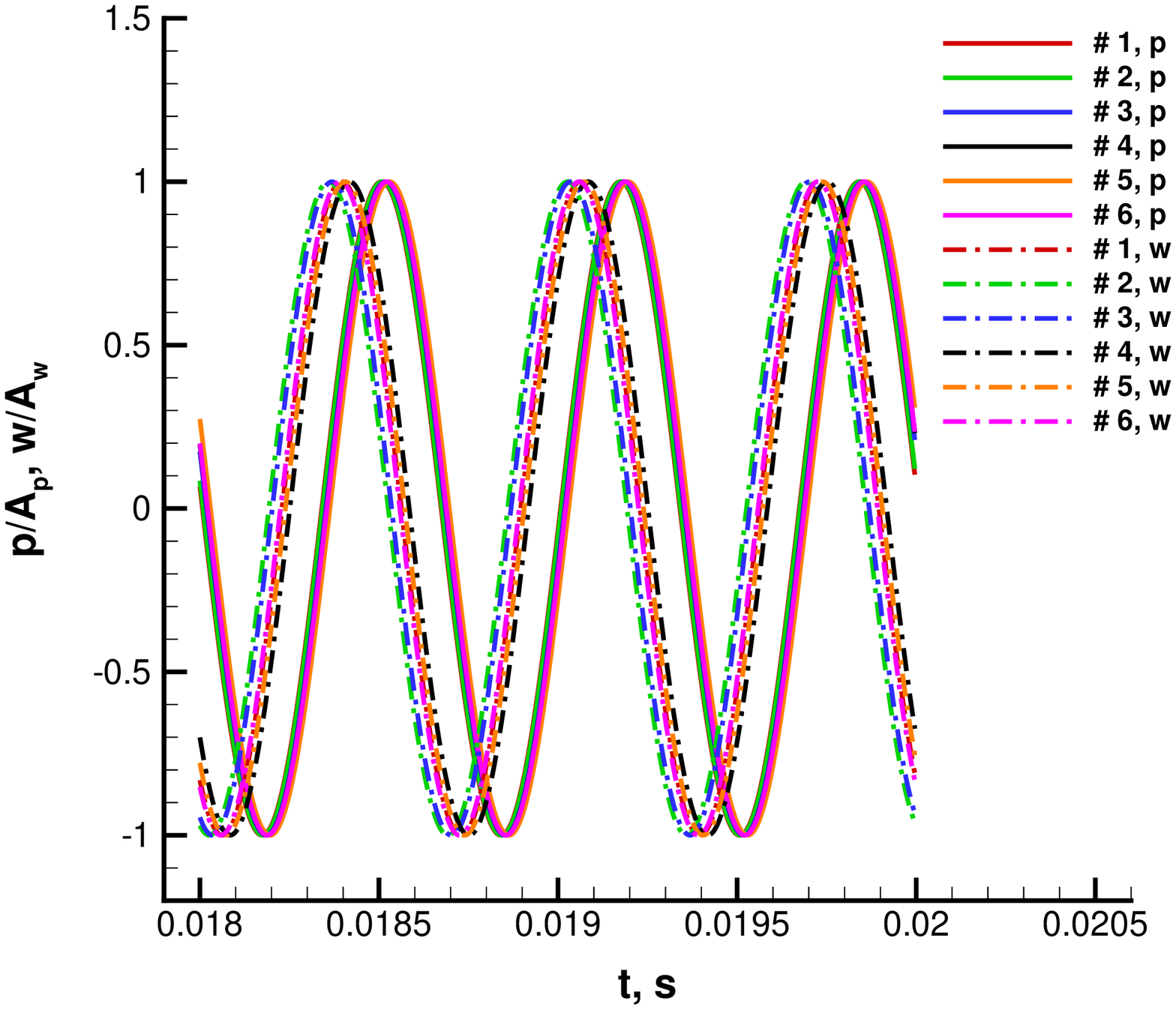}
        \label{fig:p_w_1500hz_0dot33m_osk}}
    \end{subfigmatrix}
    \caption{Rescaled time histories of pressure and axial velocity for the mode of $1500$ Hz at $z = 0.33$ m}
    \label{fig:p_w_1500hz_0dot33m}
\end{figure}

Figure~\ref{fig:p_w_1500hz_0dot01m}, \ref{fig:p_w_1500hz_0dot18m} and \ref{fig:p_w_1500hz_0dot33m} show the time histories of nondimensionalized pressure $p$ and axial velocity $w$ at $z = 0.01$ m, $z = 0.18$ m and $z = 0.33$ m for both approaches. For clear demonstration of the phase angle difference between curves, the time histories are rebuilt by only retaining the dominant mode of $1500$ Hz. The local amplitudes of the retained mode ($A_{p}$ for pressure and $A_{w}$ for axial velocity) are used as the references for nondimensionalization. $A_{p}$, $A_{w}$ and the phase angle difference $\Delta \phi$ between the time history of pressure and that of axial velocity are listed in Table~\ref{tab:1500at0dot01}, \ref{tab:1500at0dot18} and \ref{tab:1500at0dot33} for each probe. Figure~\ref{fig:p_w_1500hz_0dot33m} clearly shows that the time histories of pressure at the six different azimuthal positions are perfectly in phase at $33$ cm downstream the injector plate. Meanwhile, the time histories of axial velocity at the six probes are almost in phase as well. At each probe, the time history of pressure and that of axial velocity are nearly $90$ degrees out of phase as shown in Table~\ref{tab:1500at0dot33}. All these features indicate that the dominant mode of $1500$ Hz is a longitudinal mode.

\begin{table}
\begin{center}
\caption{Amplitudes of pressure and axial velocity modes at $1500Hz$ and their phase angle difference at the near-wall probes located at $z = 0.01m$ }
\begin{tabular}{|c|c|c|c|c|c|c|c|} \hline
\multicolumn{7}{c}{FPV} \\ \hline
  Probe & \#1 & \#2 & \#3 & \#4 & \#5 & \#6\\ \hline
  $A_{p}, Pa$ & 198332 & 184152 & 282658 & 414628 & 399987 & 362522 \\ \hline
  $A_{w}, m/s$ & 0.32 & 3.74 & 3.29 & 0.66 & 1.47 & 1.39 \\ \hline
  $\Delta \phi, ^{\circ}$ & 140 & 88 & 168 & 122 & 16 & 163 \\ \hline \hline
\multicolumn{7}{c}{one step kinetics} \\ \hline
  Probe & \#1 & \#2 & \#3 & \#4 & \#5 & \#6\\ \hline
  $A_{p}, Pa$ & 369534 & 475026 & 478063 & 324914 & 201074 & 278087 \\ \hline
  $A_{w}, m/s$ & 1.46 & 3.05 & 2.22 & 3.16 & 3.72 & 3.05 \\ \hline
  $\Delta \phi, ^{\circ}$ & 59 & 126 & 120 & 28 & 52 & 138 \\ \hline
\end{tabular}
\label{tab:1500at0dot01}
\end{center}
\end{table}

\begin{table}
\begin{center}
\caption{Amplitudes of pressure and axial velocity modes at $1500Hz$ and their phase angle difference at the near-wall probes located at $z = 0.18m$ }
\begin{tabular}{|c|c|c|c|c|c|c|c|} \hline
\multicolumn{7}{c}{FPV} \\ \hline
  Probe & \#1 & \#2 & \#3 & \#4 & \#5 & \#6\\ \hline
  $A_{p}, Pa$ & 36184 & 46793 & 73922 & 89115 & 102933 & 82449 \\ \hline
  $A_{w}, m/s$ & 21.26 & 20.86 & 19.28 & 20.99 & 17.95 & 15.09 \\ \hline
  $\Delta \phi, ^{\circ}$ & 55 & 13 & 43 & 75 & 86 & 94 \\ \hline \hline
\multicolumn{7}{c}{one step kinetics} \\ \hline
  Probe & \#1 & \#2 & \#3 & \#4 & \#5 & \#6\\ \hline
  $A_{p}, Pa$ & 118531 & 166317 & 157082 & 94782 & 40907 & 49387 \\ \hline
  $A_{w}, m/s$ & 18.4 & 20.11 & 22.54 & 15.42 & 14.24 & 17.64 \\ \hline
  $\Delta \phi, ^{\circ}$ & 61 & 76 & 76 & 72 & 79 & 91 \\ \hline
\end{tabular}
\label{tab:1500at0dot18}
\end{center}
\end{table}

\begin{table}
\begin{center}
\caption{Amplitudes of pressure and axial velocity modes at $1500Hz$ and their phase angle difference at the near-wall probes located at $z = 0.33m$ }
\begin{tabular}{|c|c|c|c|c|c|c|c|} \hline
\multicolumn{7}{c}{FPV} \\ \hline
  Probe & \#1 & \#2 & \#3 & \#4 & \#5 & \#6\\ \hline
  $A_{p}, Pa$ & 240412 & 221046 & 239366 & 274673 & 284333 & 254738 \\ \hline
  $A_{w}, m/s$ & 8.73 & 10.6 & 9.19 & 5.33 & 11.86 & 11.86 \\ \hline
  $\Delta \phi, ^{\circ}$ & 79 & 78 & 88 & 87 & 85 & 92 \\ \hline \hline
\multicolumn{7}{c}{one step kinetics} \\ \hline
  Probe & \#1 & \#2 & \#3 & \#4 & \#5 & \#6\\ \hline
  $A_{p}, Pa$ & 241048 & 235499 & 231516 & 240684 & 258771 & 262803 \\ \hline
  $A_{w}, m/s$ & 8.8 & 9.87 & 10.0 & 13.27 & 11.62 & 8.25 \\ \hline
  $\Delta \phi, ^{\circ}$ & 60 & 81 & 81 & 55 & 67 & 69 \\ \hline
\end{tabular}
\label{tab:1500at0dot33}
\end{center}
\end{table}

\begin{figure}[H]
    \begin{subfigmatrix}{2}
        \subfigure[FPV]
{\includegraphics{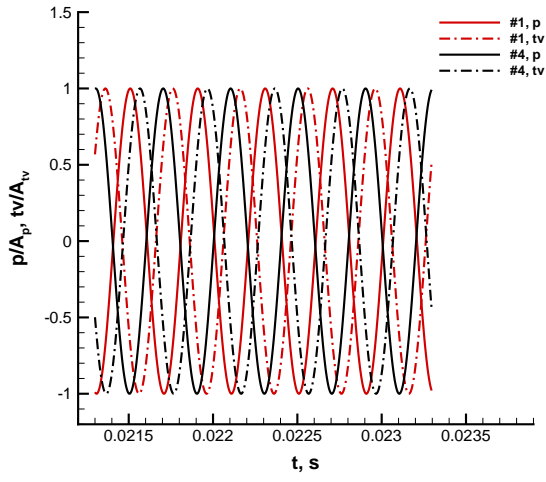}
        \label{fig:p_tv_2500hz_0dot01m_fpv_probe_1_and_4}}
        \subfigure[one step kinetics]
{\includegraphics{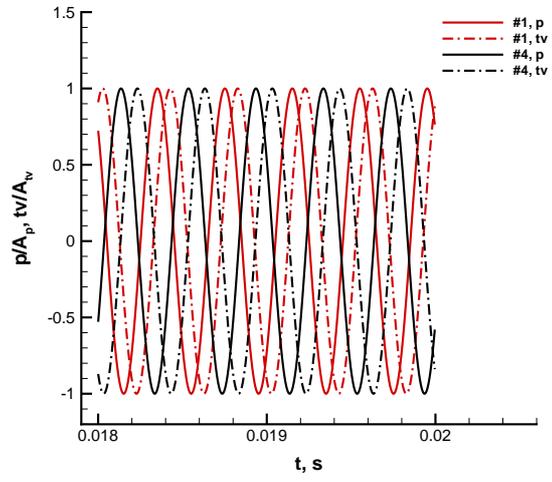}
        \label{fig:p_tv_2500hz_0dot01m_osk_probe_1_and_4}}
    \end{subfigmatrix}
    \caption{Rescaled time histories of pressure and tangential velocity at probes \# 1 and \# 4 for the mode of $2500$ Hz at $z = 0.01$ m}
    \label{fig:p_tv_2500hz_0dot01m_probe_1_and_4}
\end{figure}

\begin{figure}[H]
    \begin{subfigmatrix}{2}
        \subfigure[FPV]
{\includegraphics{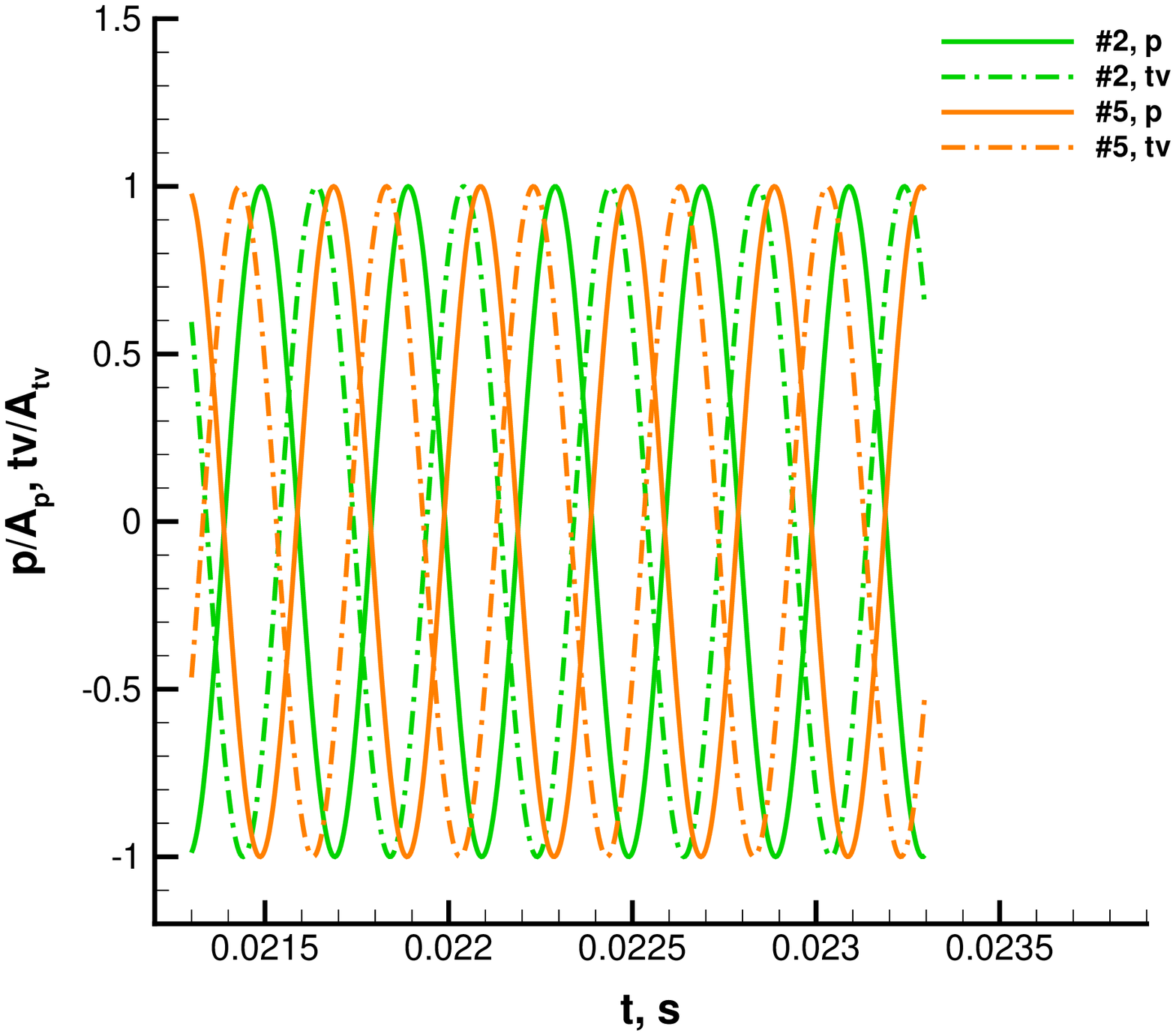}
        \label{fig:p_tv_2500hz_0dot01m_fpv_probe_2_and_5}}
        \subfigure[one step kinetics]
{\includegraphics{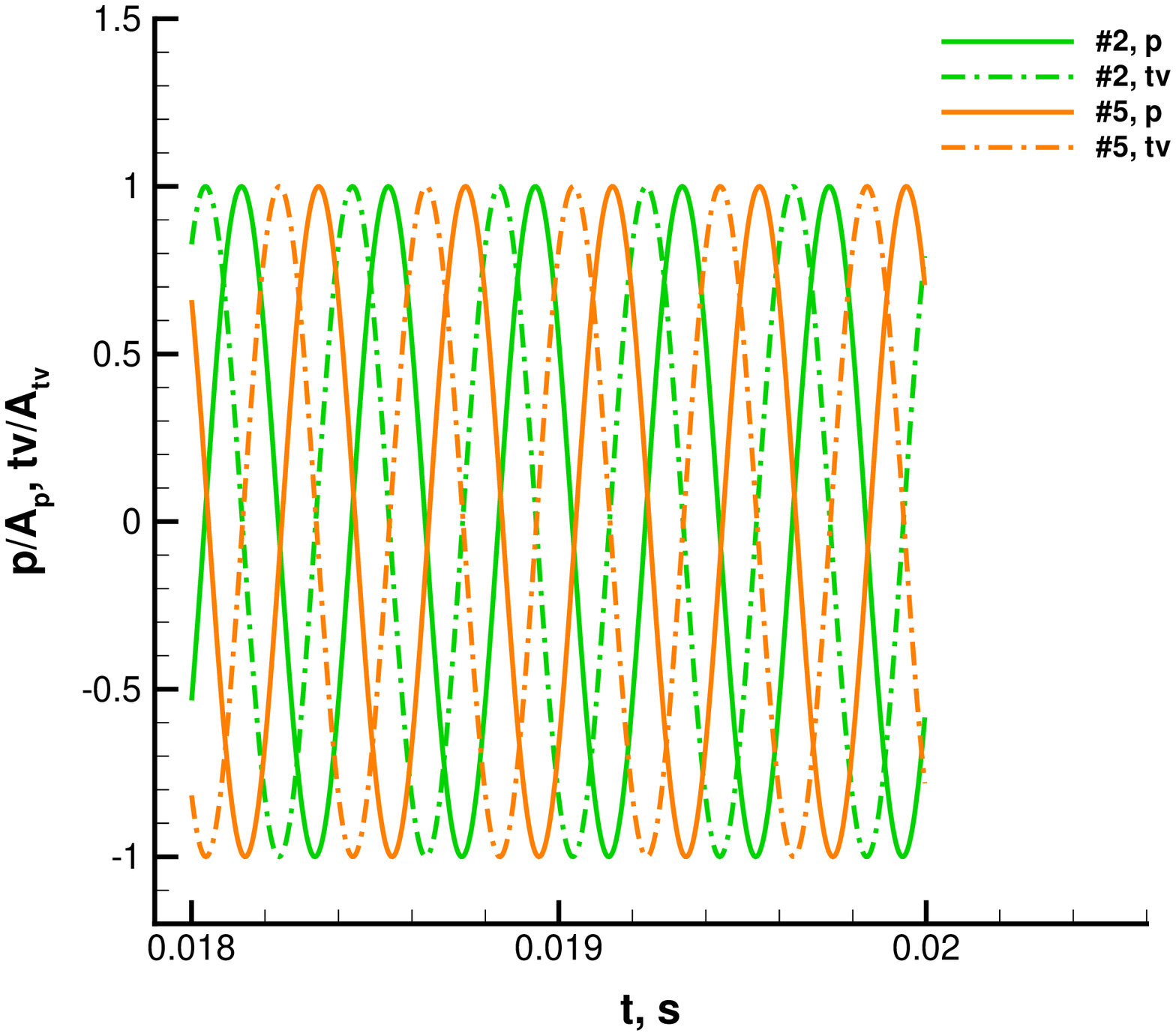}
        \label{fig:p_tv_2500hz_0dot01m_osk_probe_2_and_5}}
    \end{subfigmatrix}
    \caption{Rescaled time histories of pressure and tangential velocity at probes \# 2 and \# 5 for the mode of $2500$ Hz at $z = 0.01$ m}
    \label{fig:p_tv_2500hz_0dot01m_probe_2_and_5}
\end{figure}

\begin{figure}[H]
    \begin{subfigmatrix}{2}
        \subfigure[FPV]
{\includegraphics{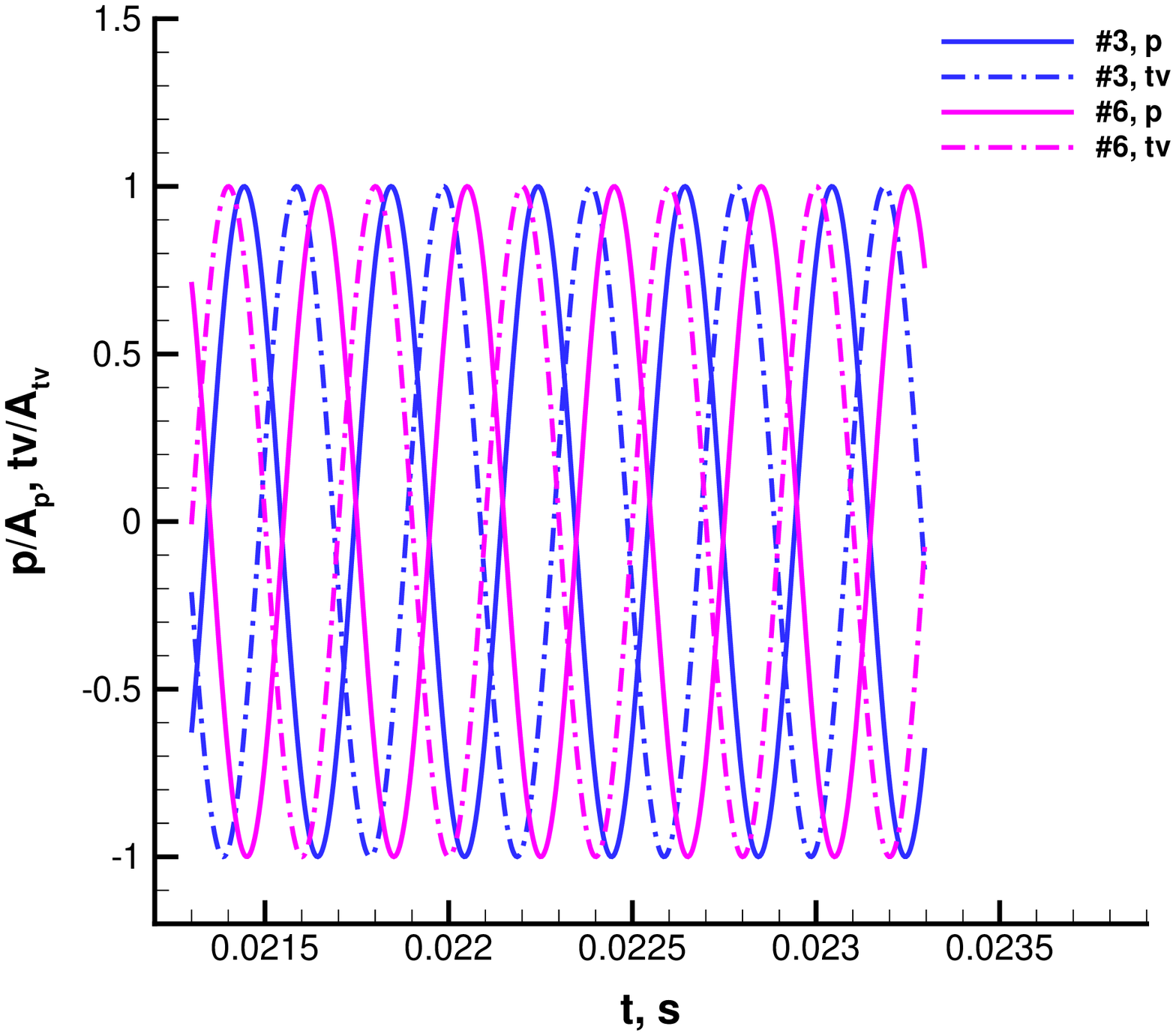}
        \label{fig:p_tv_2500hz_0dot01m_fpv_probe_3_and_6}}
        \subfigure[one step kinetics]
{\includegraphics{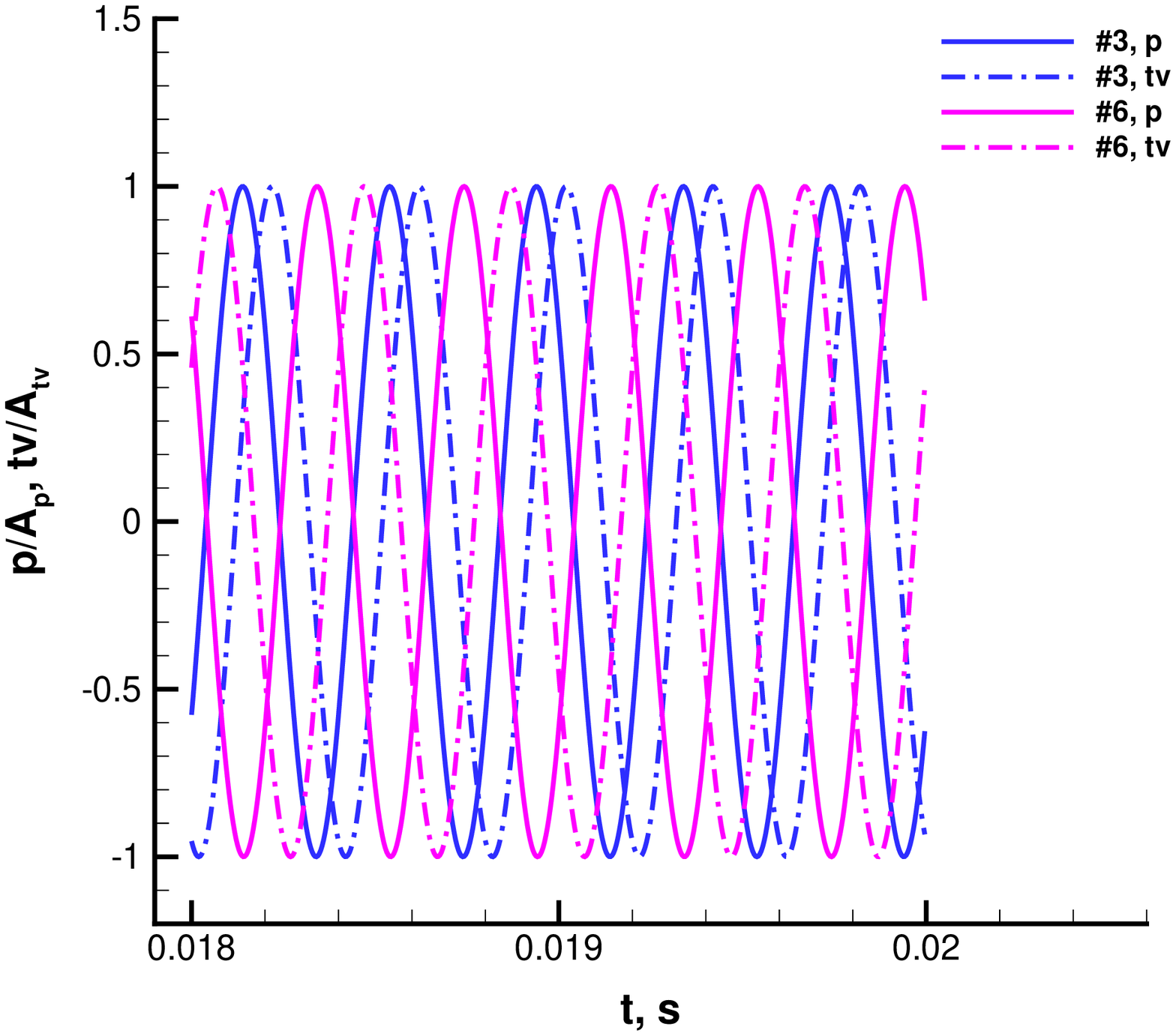}
        \label{fig:p_tv_2500hz_0dot01m_osk_probe_3_and_6}}
    \end{subfigmatrix}
    \caption{Rescaled time histories of pressure and tangential velocity at probes \# 3 and \# 6 for the mode of $2500$ Hz at $z = 0.01$ m}
    \label{fig:p_tv_2500hz_0dot01m_probe_3_and_6}
\end{figure}

\begin{figure}[H]
    \begin{subfigmatrix}{2}
        \subfigure[FPV]
{\includegraphics{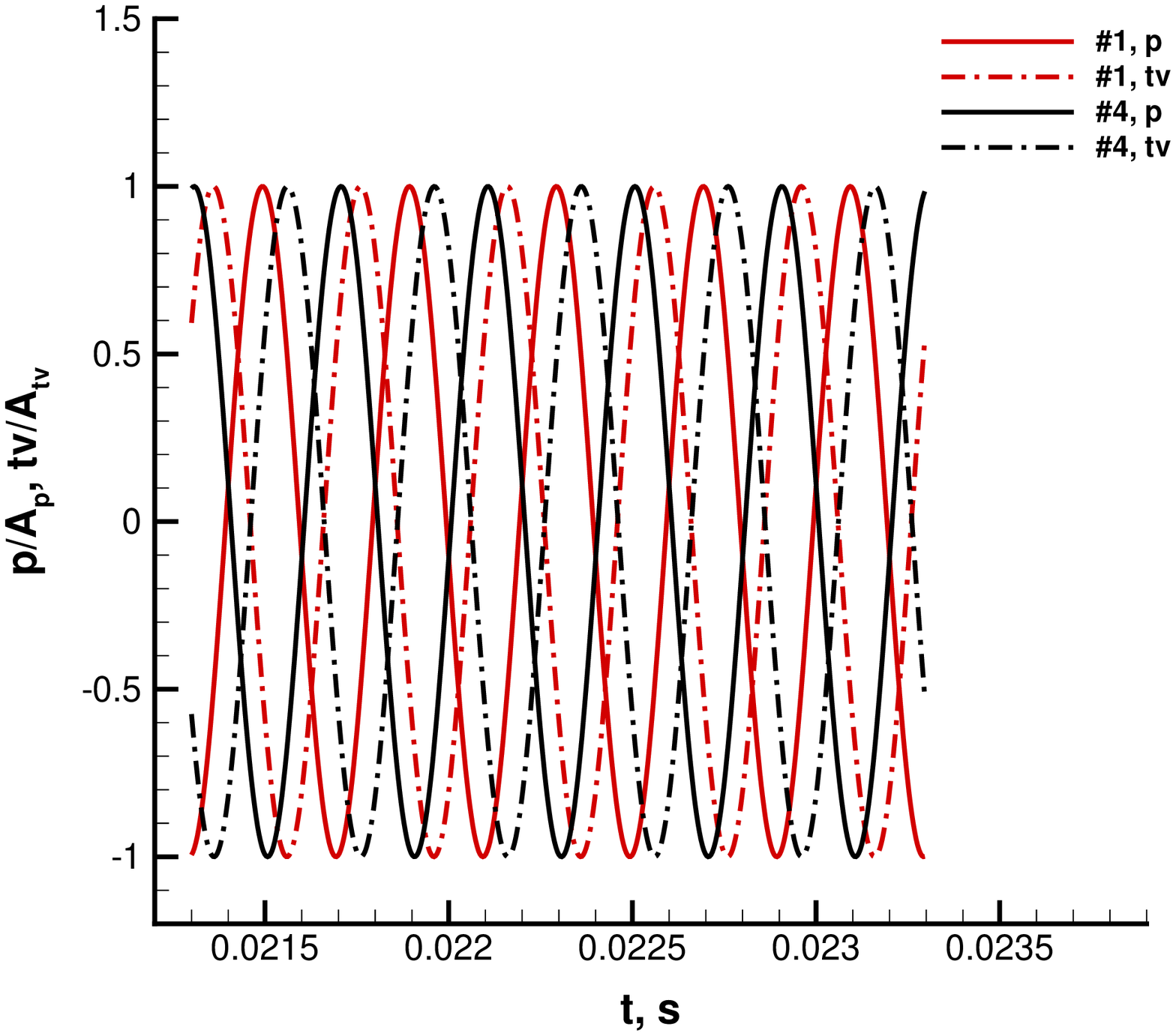}
        \label{fig:p_tv_2500hz_0dot18m_fpv_probe_1_and_4}}
        \subfigure[one step kinetics]
{\includegraphics{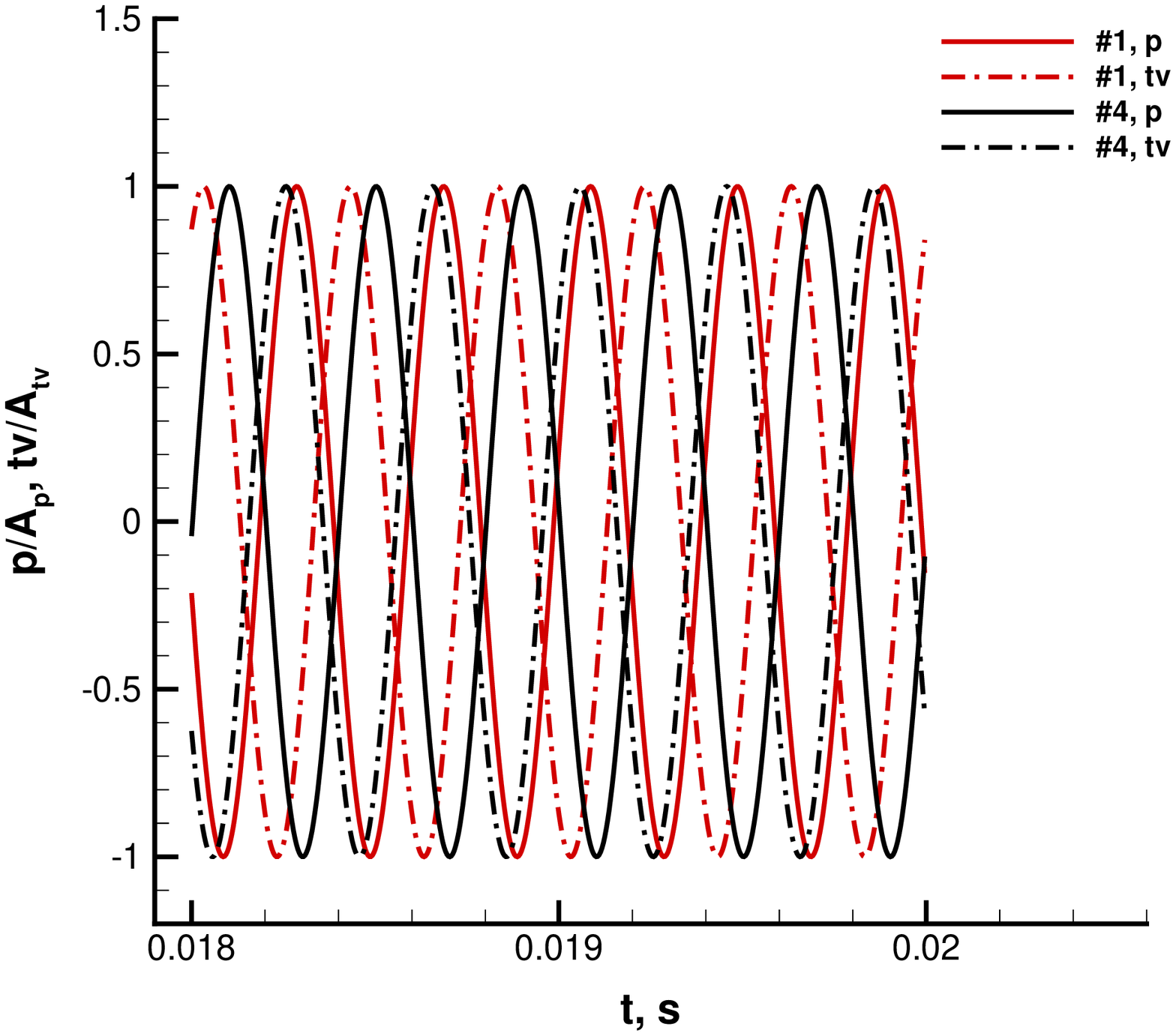}
        \label{fig:p_tv_2500hz_0dot18m_osk_probe_1_and_4}}
    \end{subfigmatrix}
    \caption{Rescaled time histories of pressure and tangential velocity at probes \# 1 and \# 4 for the mode of $2500$ Hz at $z = 0.18$ m}
    \label{fig:p_tv_2500hz_0dot18m_probe_1_and_4}
\end{figure}

\begin{figure}[H]
    \begin{subfigmatrix}{2}
        \subfigure[FPV]
{\includegraphics{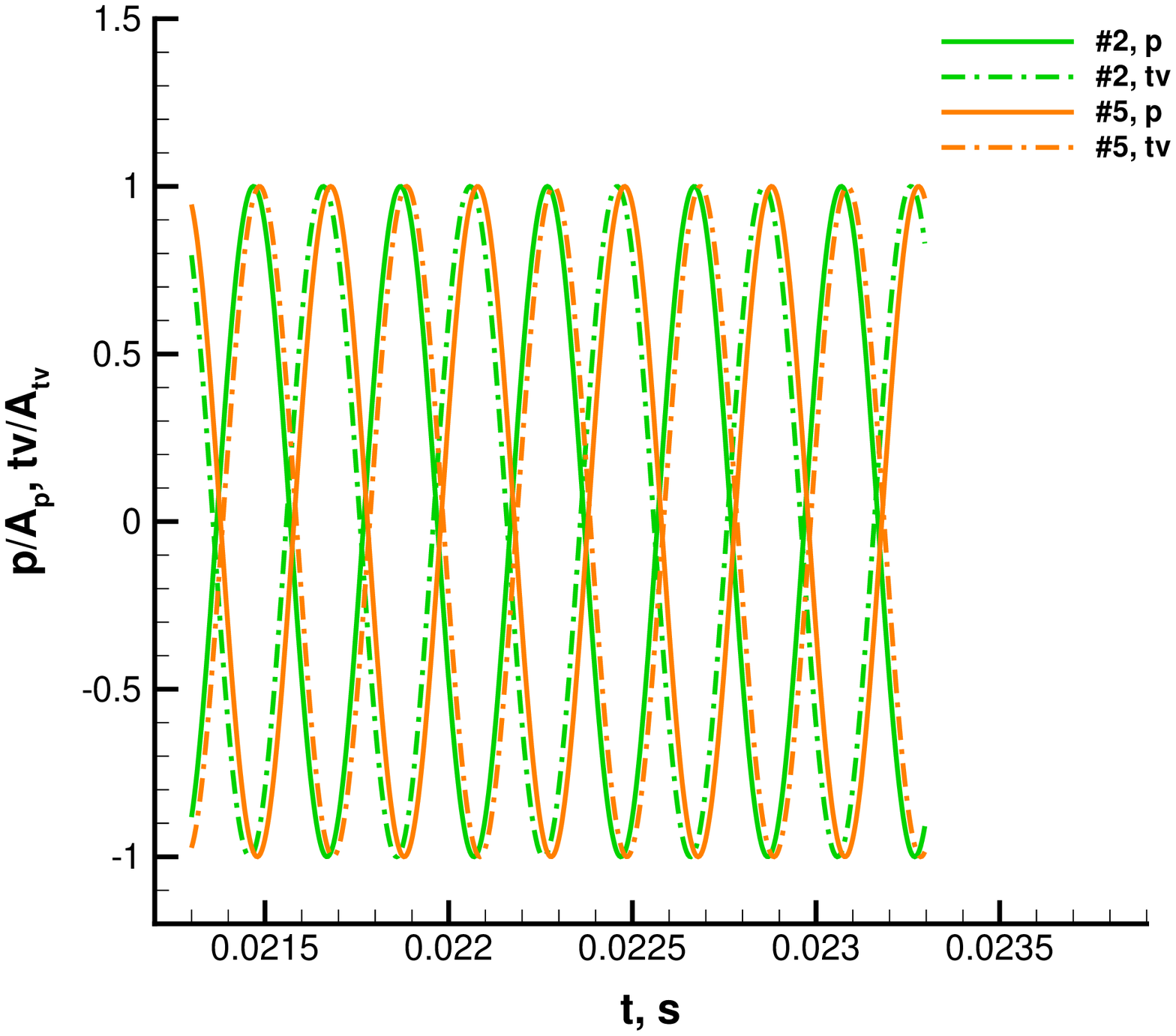}
        \label{fig:p_tv_2500hz_0dot18m_fpv_probe_2_and_5}}
        \subfigure[one step kinetics]
{\includegraphics{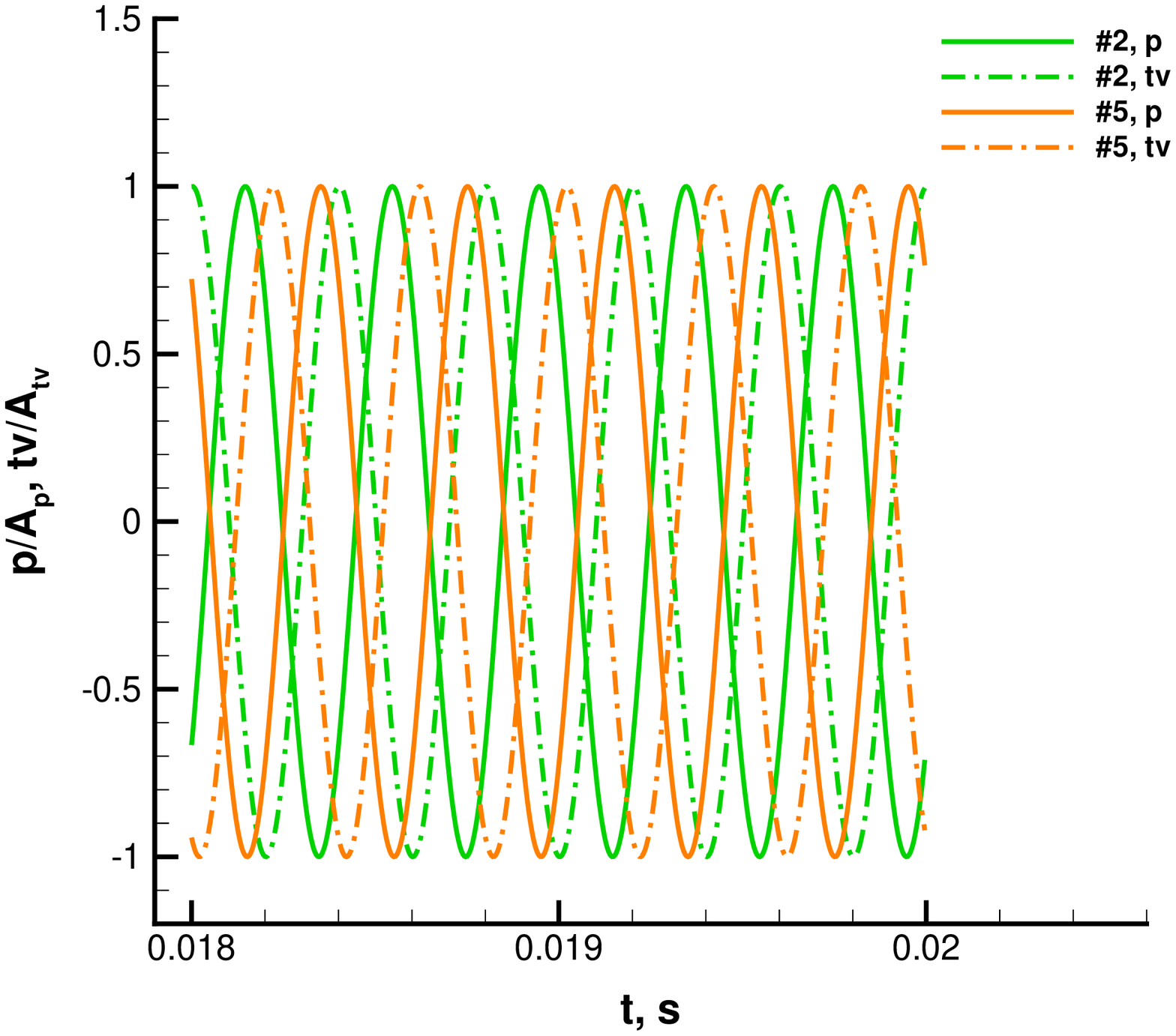}
        \label{fig:p_tv_2500hz_0dot18m_osk_probe_2_and_5}}
    \end{subfigmatrix}
    \caption{Rescaled time histories of pressure and tangential velocity at probes \# 2 and \# 5 for the mode of $2500$ Hz at $z = 0.18$ m}
    \label{fig:p_tv_2500hz_0dot18m_probe_2_and_5}
\end{figure}

\begin{figure}[H]
    \begin{subfigmatrix}{2}
        \subfigure[FPV]
{\includegraphics{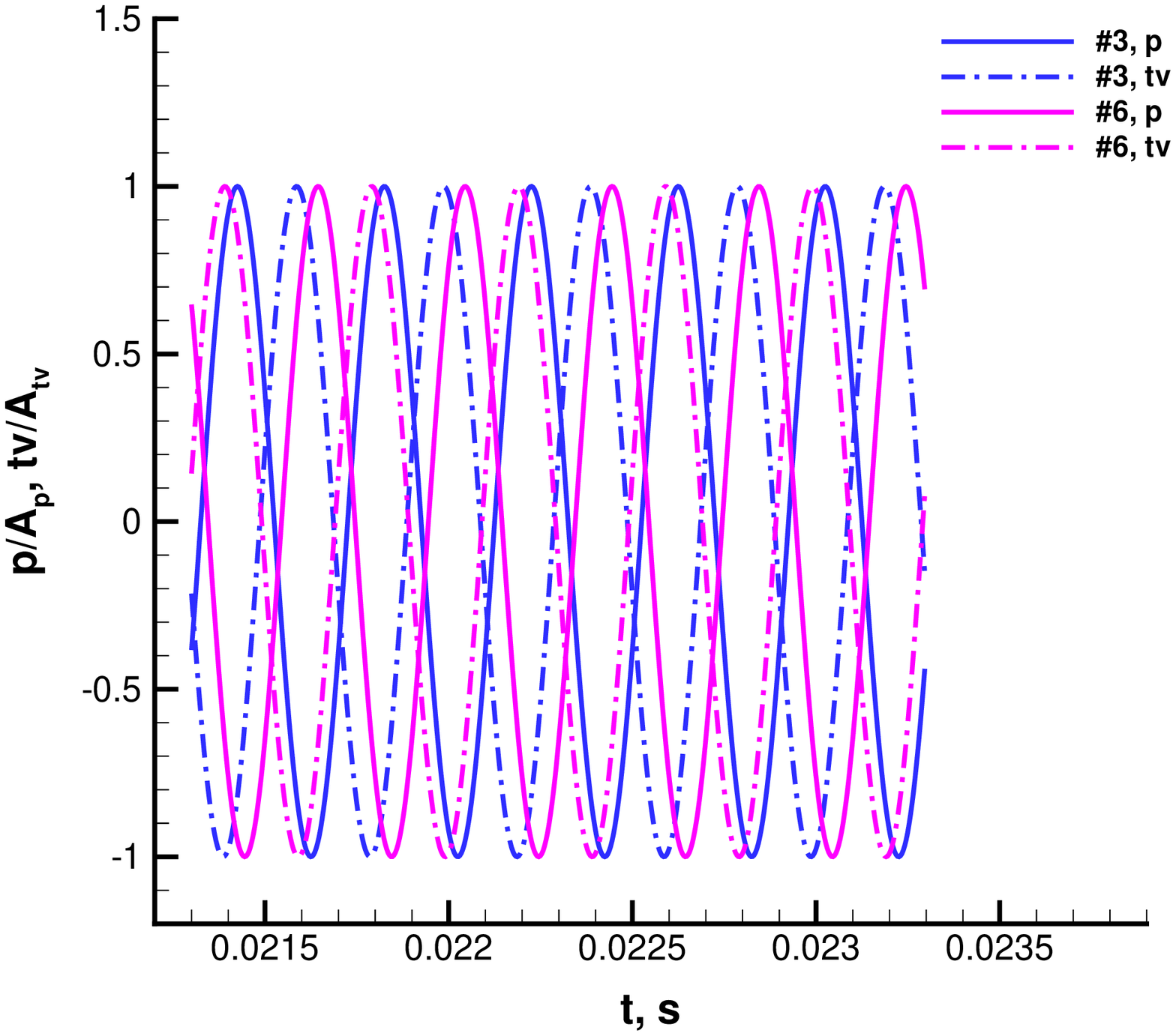}
        \label{fig:p_tv_2500hz_0dot18m_fpv_probe_3_and_6}}
        \subfigure[one step kinetics]
{\includegraphics{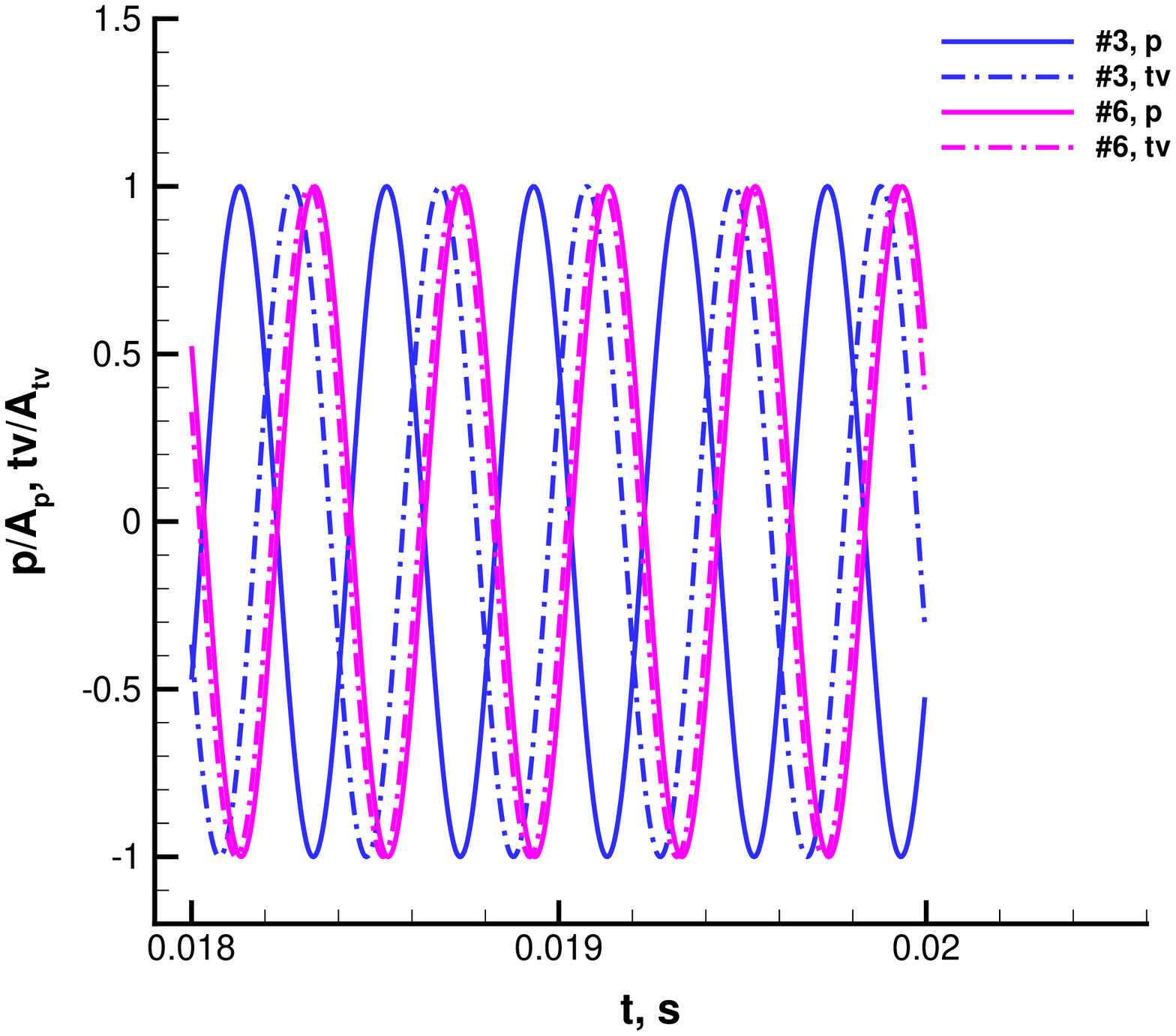}
        \label{fig:p_tv_2500hz_0dot18m_osk_probe_3_and_6}}
    \end{subfigmatrix}
    \caption{Rescaled time histories of pressure and tangential velocity at probes \# 3 and \# 6 for the mode of $2500$ Hz at $z = 0.18$ m}
    \label{fig:p_tv_2500hz_0dot18m_probe_3_and_6}
\end{figure}

\begin{figure}[H]
    \begin{subfigmatrix}{2}
        \subfigure[FPV]
{\includegraphics{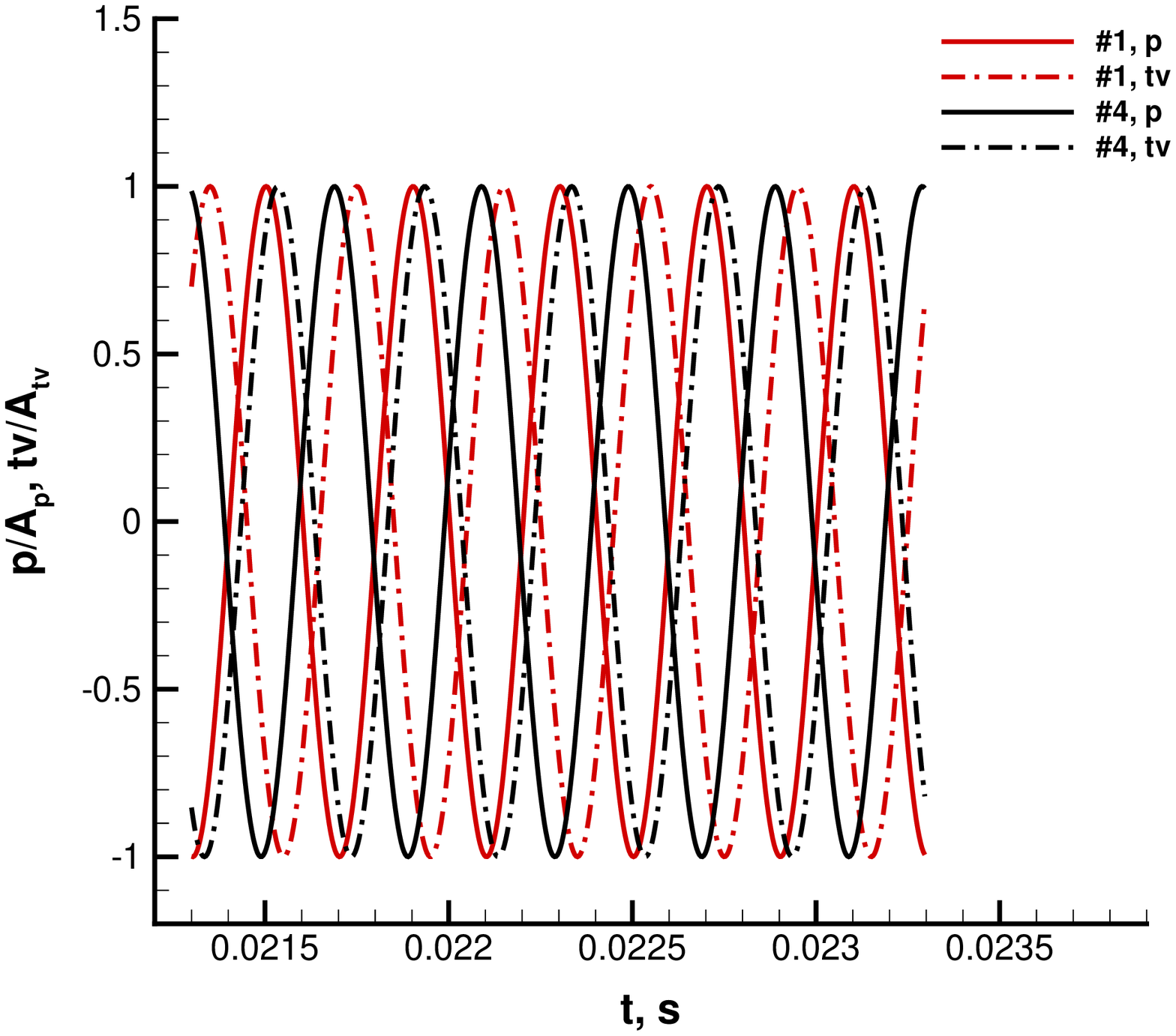}
        \label{fig:p_tv_2500hz_0dot33m_fpv_probe_1_and_4}}
        \subfigure[one step kinetics]
{\includegraphics{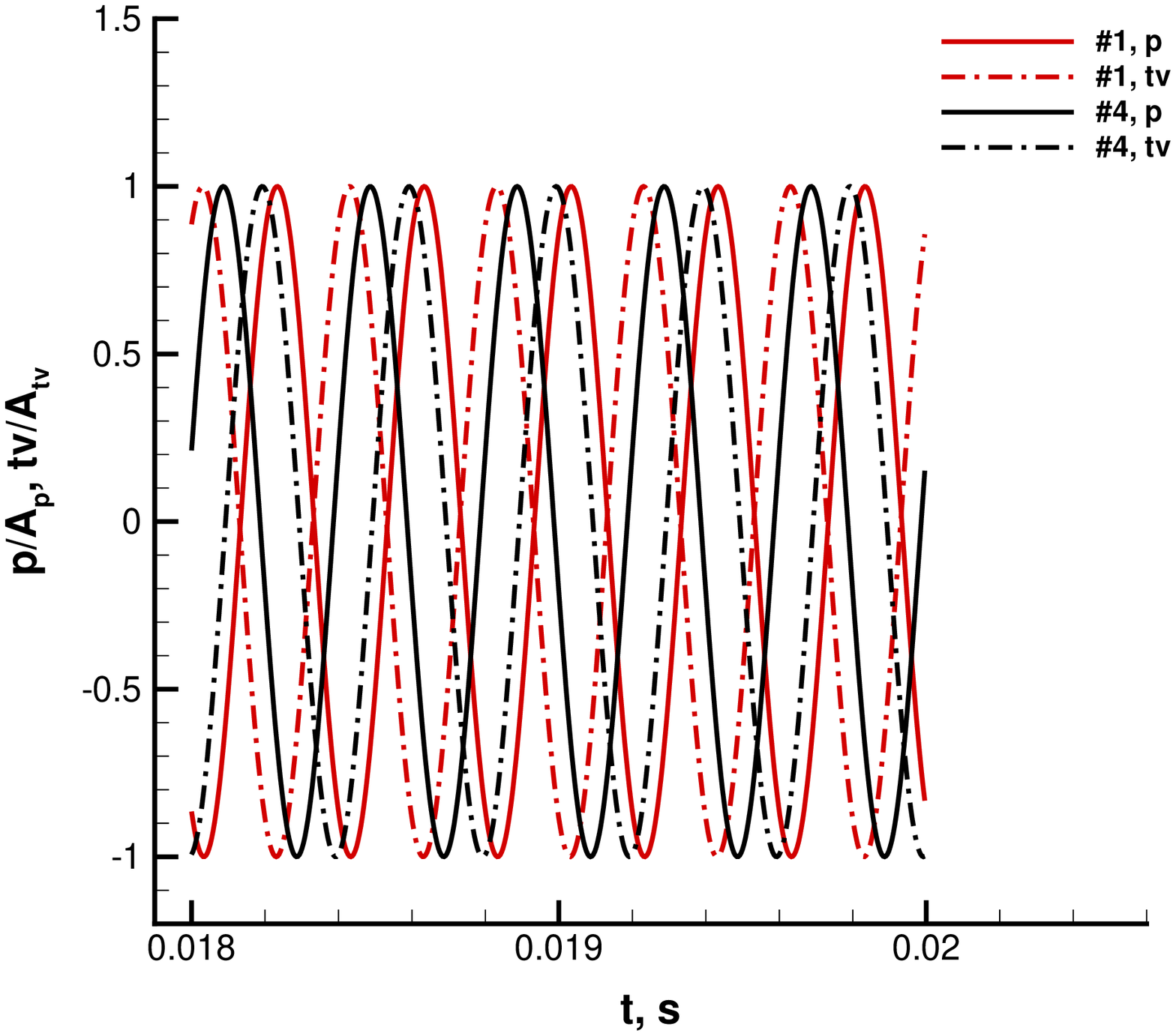}
        \label{fig:p_tv_2500hz_0dot33m_osk_probe_1_and_4}}
    \end{subfigmatrix}
    \caption{Rescaled time histories of pressure and tangential velocity at probes \# 1 and \# 4 for the mode of $2500$ Hz at $z = 0.33$ m}
    \label{fig:p_tv_2500hz_0dot33m_probe_1_and_4}
\end{figure}

\begin{figure}[H]
    \begin{subfigmatrix}{2}
        \subfigure[FPV]
{\includegraphics{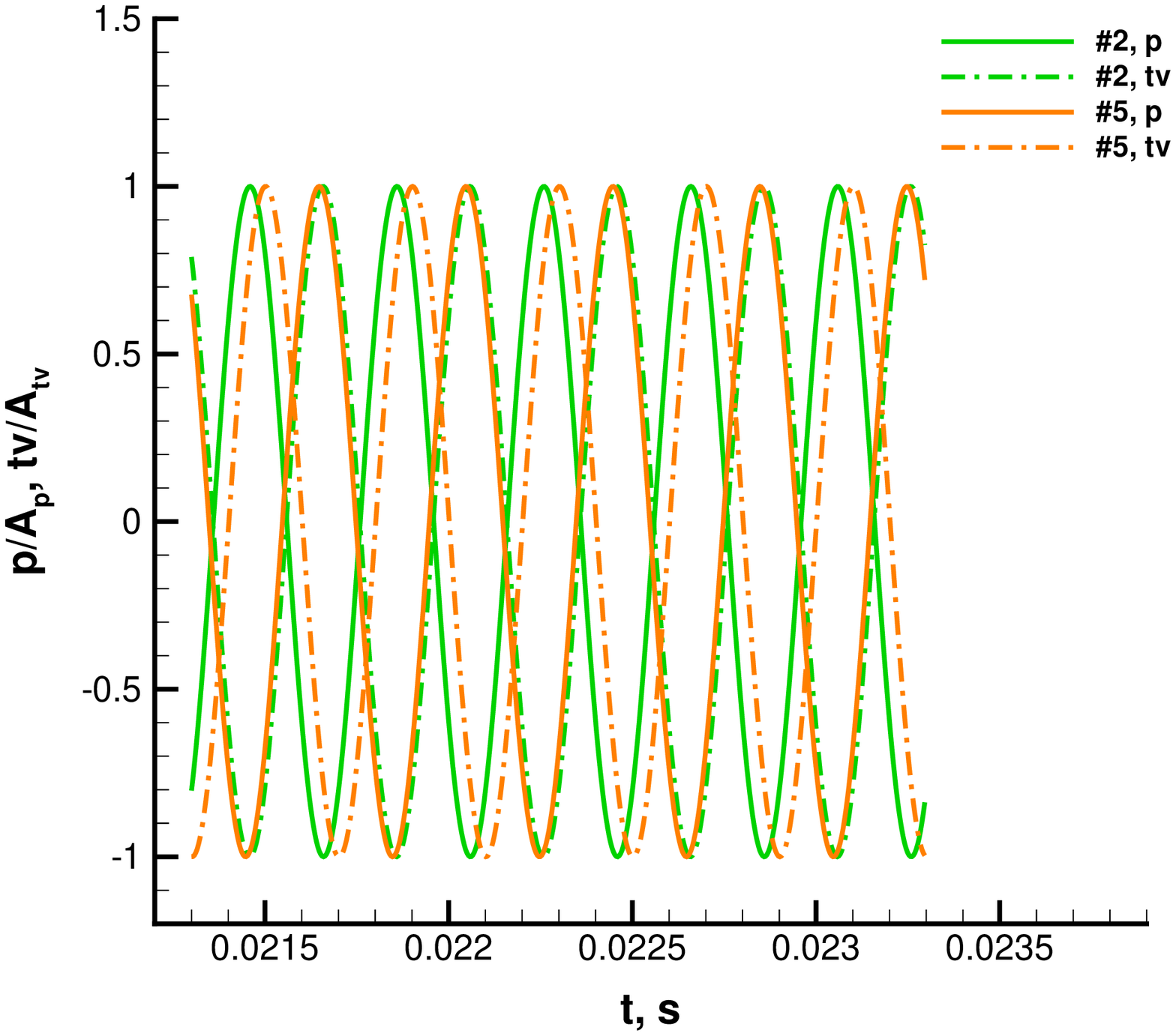}
        \label{fig:p_tv_2500hz_0dot33m_fpv_probe_2_and_5}}
        \subfigure[one step kinetics]
{\includegraphics{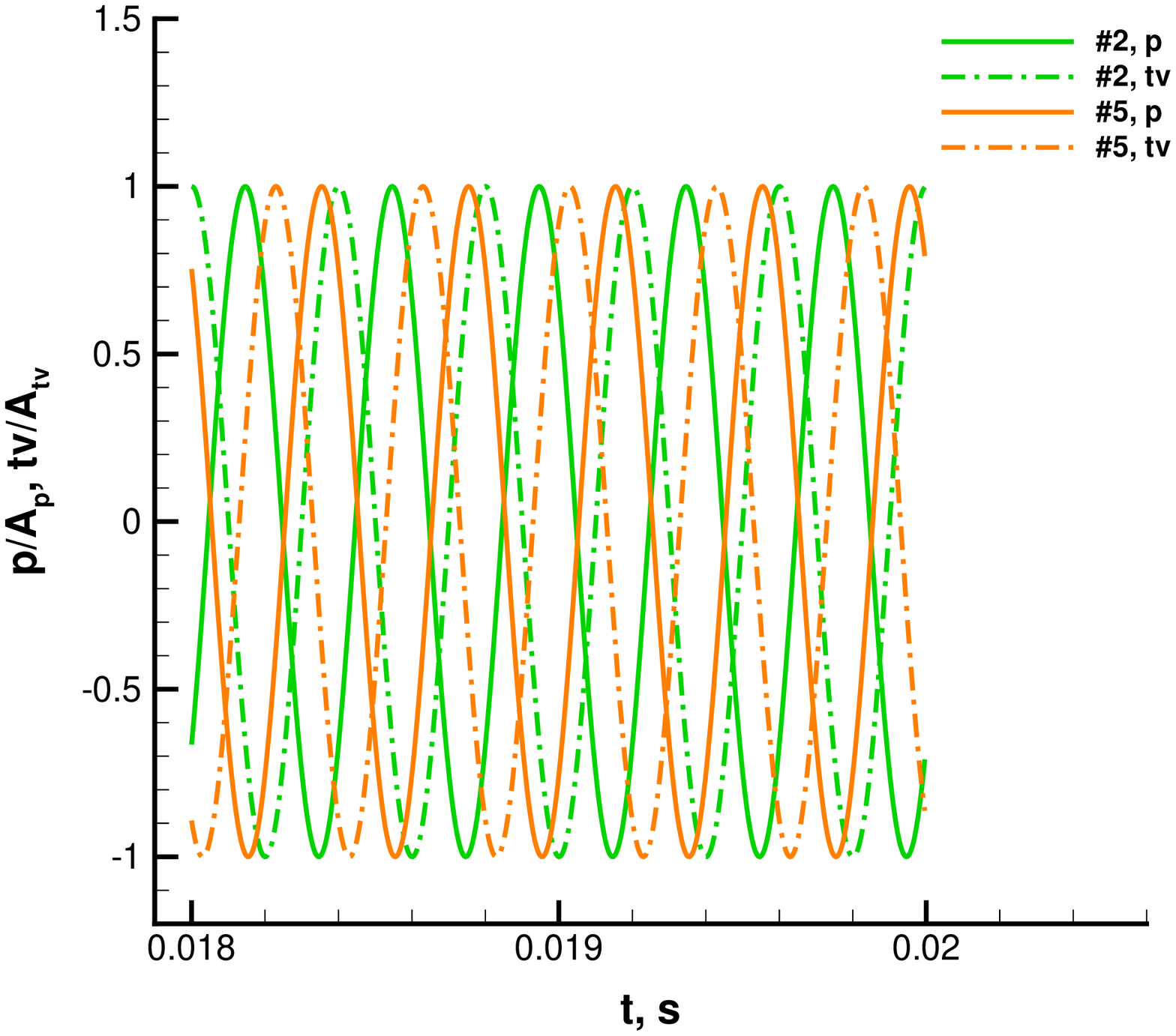}
        \label{fig:p_tv_2500hz_0dot33m_osk_probe_2_and_5}}
    \end{subfigmatrix}
    \caption{Rescaled time histories of pressure and tangential velocity at probes \# 2 and \# 5 for the mode of $2500$ Hz at $z = 0.33$ m}
    \label{fig:p_tv_2500hz_0dot33m_probe_2_and_5}
\end{figure}

\begin{figure}[H]
    \begin{subfigmatrix}{2}
        \subfigure[FPV]
{\includegraphics{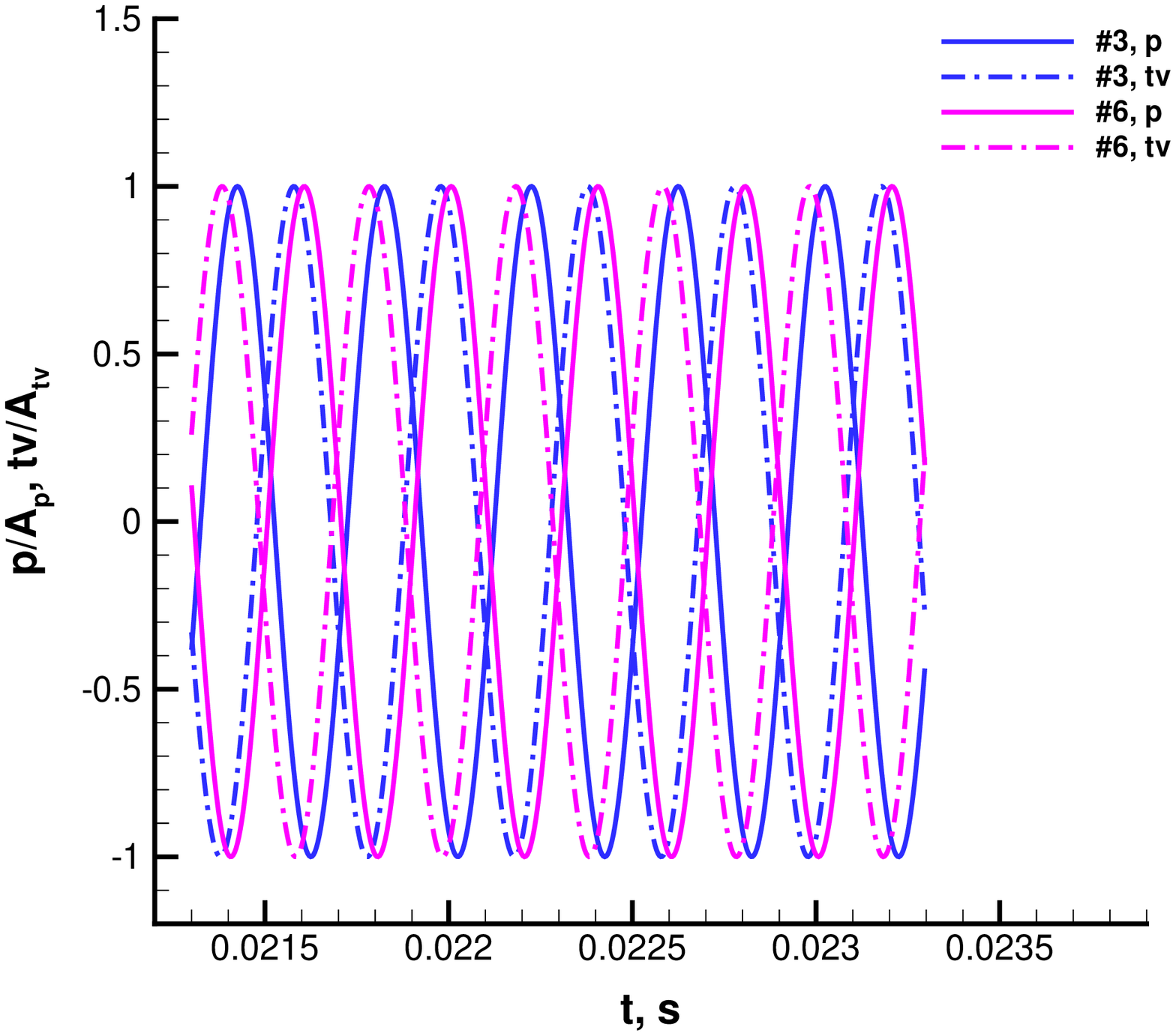}
        \label{fig:p_tv_2500hz_0dot33m_fpv_probe_3_and_6}}
        \subfigure[one step kinetics]
{\includegraphics{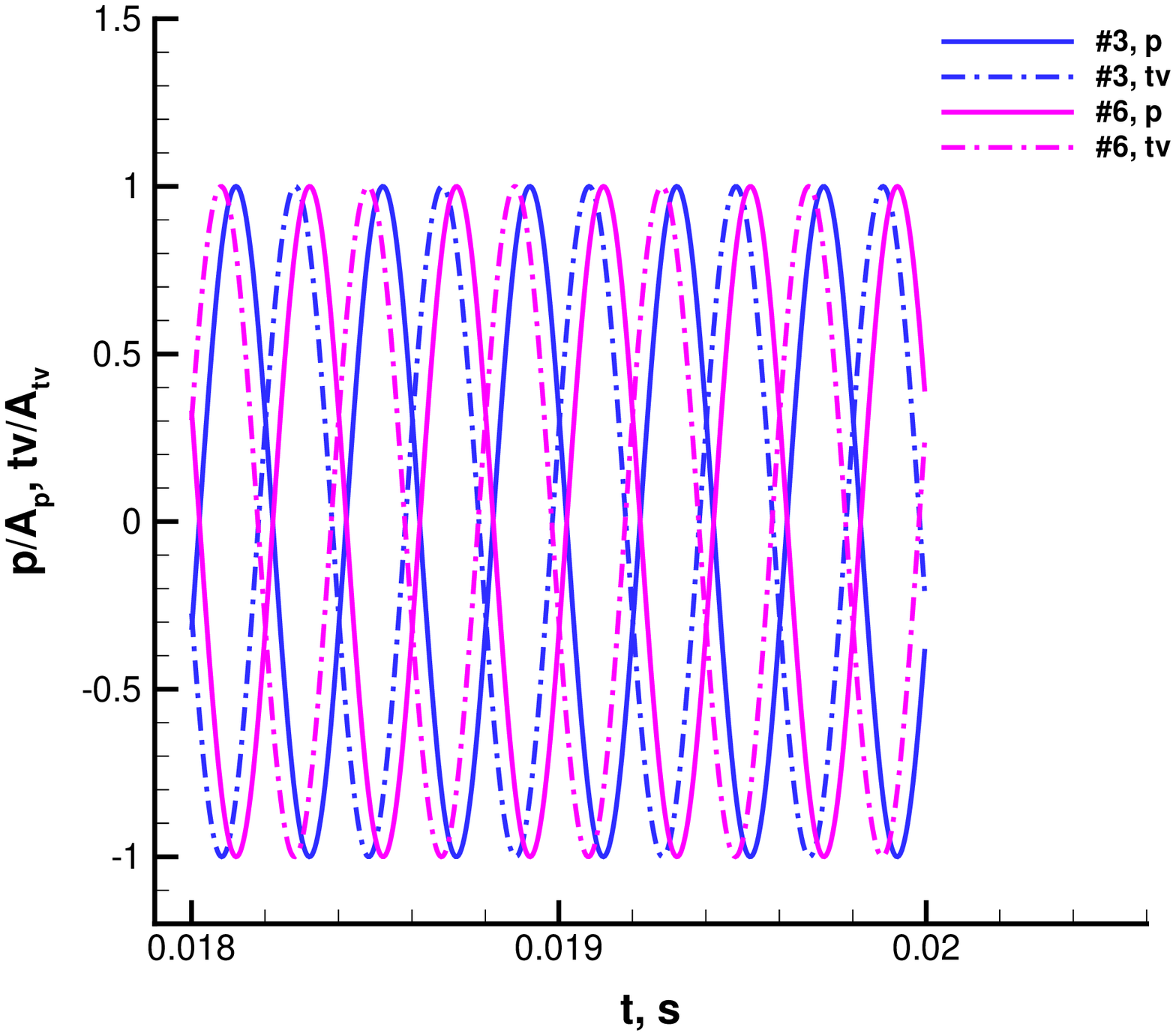}
        \label{fig:p_tv_2500hz_0dot33m_osk_probe_3_and_6}}
    \end{subfigmatrix}
    \caption{Rescaled time histories of pressure and tangential velocity at probes \# 3 and \# 6 for the mode of $2500$ Hz at $z = 0.33$ m}
    \label{fig:p_tv_2500hz_0dot33m_probe_3_and_6}
\end{figure}

Figure~\ref{fig:p_tv_2500hz_0dot01m_probe_1_and_4} to \ref{fig:p_tv_2500hz_0dot33m_probe_3_and_6} show the time histories of nondimensional pressure and tangential velocity at $z = 0.01$ m, $z = 0.18$ m and $z = 0.33$ m for both approaches. Similar to the above analysis of the longitudinal mode, the time histories are rebuilt by only retaining the dominant mode of $2500$ Hz. The amplitudes of the retained mode ($A_{p}$ for pressure and $A_{tv}$ for tangential velocity) are used as the reference for nondimensionalization. $A_{p}$, $A_{tv}$ and the phase angle difference $\Delta \phi$ between the time history of pressure and that of tangential velocity are listed in Table~\ref{tab:2500at0dot01}, \ref{tab:2500at0dot18} and \ref{tab:2500at0dot33} for each probe. 

Figure~\ref{fig:p_tv_2500hz_0dot01m_probe_1_and_4} to \ref{fig:p_tv_2500hz_0dot33m_probe_3_and_6} show that, at each pair of probes that are $180$ degrees away from each other in the circumferential direction, the time histories of pressure $p$ are almost $180$ degrees out of phase. So are the time histories of tangential velocity. These features are observed for both the FPV and the OSK approaches. As listed in Table~\ref{tab:2500at0dot01}, \ref{tab:2500at0dot18} and \ref{tab:2500at0dot33}, the amplitude of the pressure mode of $2500$ Hz achieves maximum on probes \#2/\#5 and minimum on probes \#3/\#6 for the FPV approach. For the OSK approach, the amplitude of the same pressure mode achieves maximum on probes \#3/\#6 and minimum on probes \#1/\#4. For the tangential mode of $2500$ Hz, its amplitude gets maximum/minimum where the amplitude of pressure mode reaches minimum/maximum. This is verified for both the FPV and OSK approaches. This information suggests pressure antinode (or node of tangential velocity) near probes \#2/\#5 and pressure node (or antinode of tangential velocity) near probes \#3/\#6 for the FPV approach. For the OSK approach, the pressure antinode and node are near probes \#3/\#6 and \#1/\#4, respectively. Hence, the dominant mode of $2500$ Hz is a tangential standing wave. 

On each probe, the time history of pressure $p$ and that of tangential velocity $tv$ should be $90$ degrees out of phase for a tangential standing wave. However, $\Delta \phi$ listed in Table~\ref{tab:2500at0dot01}, \ref{tab:2500at0dot18} and \ref{tab:2500at0dot33} deviates remarkably from the theoretical value. A possible reason is that the probes are placed relatively close to the chamber wall where boundary layer can affect the velocity phase.

\begin{table}
\begin{center}
\caption{Amplitudes of pressure and tangential velocity modes at $2500Hz$ and their phase angle difference at the near-wall probes located at $z = 0.01m$ }
\begin{tabular}{|c|c|c|c|c|c|c|c|} \hline
\multicolumn{7}{c}{FPV} \\ \hline
  Probe & \#1 & \#2 & \#3 & \#4 & \#5 & \#6\\ \hline
  $A_{p}, Pa$ & 597535 & 816468 & 304913 & 633325 & 807362 & 333843 \\ \hline
  $A_{tv}, m/s$ & 17.84 & 9.92 & 25.35 & 21.07 & 9.02 & 26.29 \\ \hline
  $\Delta \phi, ^{\circ}$ & 135 & 135 & 127 & 124 & 130 & 134 \\ \hline \hline
\multicolumn{7}{c}{one step kinetics} \\ \hline
  Probe & \#1 & \#2 & \#3 & \#4 & \#5 & \#6\\ \hline
  $A_{p}, Pa$ & 130366 & 266461 & 405821 & 130586 & 269385 & 401360 \\ \hline
  $A_{tv}, m/s$ & 9.22 & 8.96 & 1.97 & 9.86 & 9.35 & 3.21 \\ \hline
  $\Delta \phi, ^{\circ}$ & 69 & 88 & 72 & 88 & 96 & 115 \\ \hline
\end{tabular}
\label{tab:2500at0dot01}
\end{center}
\end{table}

\begin{table}
\begin{center}
\caption{Amplitudes of pressure and tangential velocity modes at $2500Hz$ and their phase angle difference at the near-wall probes located at $z = 0.18m$ }
\begin{tabular}{|c|c|c|c|c|c|c|c|} \hline
\multicolumn{7}{c}{FPV} \\ \hline
  Probe & \#1 & \#2 & \#3 & \#4 & \#5 & \#6\\ \hline
  $A_{p}, Pa$ & 357370 & 560478 & 293914 & 387845 & 531930 & 245570 \\ \hline
  $A_{tv}, m/s$ & 13.24 & 5.58 & 16.69 & 14.3 & 5.85 & 14.16 \\ \hline
  $\Delta \phi, ^{\circ}$ & 120 & 171 & 145 & 131 & 174 & 131 \\ \hline \hline
\multicolumn{7}{c}{one step kinetics} \\ \hline
  Probe & \#1 & \#2 & \#3 & \#4 & \#5 & \#6\\ \hline
  $A_{p}, Pa$ & 73548 & 222285 & 292992 & 100620 & 205538 & 287141 \\ \hline
  $A_{tv}, m/s$ & 7.83 & 5.07 & 4.22 & 9.49 & 6.32 & 1.71 \\ \hline
  $\Delta \phi, ^{\circ}$ & 131 & 131 & 130 & 139 & 117 & 13 \\ \hline
\end{tabular}
\label{tab:2500at0dot18}
\end{center}
\end{table}

\begin{table}
\begin{center}
\caption{Amplitudes of pressure and tangential velocity modes at $2500Hz$ and their phase angle difference at the near-wall probes located at $z = 0.33m$ }
\begin{tabular}{|c|c|c|c|c|c|c|c|} \hline
\multicolumn{7}{c}{FPV} \\ \hline
  Probe & \#1 & \#2 & \#3 & \#4 & \#5 & \#6\\ \hline
  $A_{p}, Pa$ & 177521 & 240494 & 159256 & 112908 & 214050 & 163417 \\ \hline
  $A_{tv}, m/s$ & 8.22 & 4.17 & 8.28 & 6.94 & 3.17 & 6.51 \\ \hline
  $\Delta \phi, ^{\circ}$ & 137 & 178 & 138 & 139 & 131 & 159 \\ \hline \hline
\multicolumn{7}{c}{one step kinetics} \\ \hline
  Probe & \#1 & \#2 & \#3 & \#4 & \#5 & \#6\\ \hline
  $A_{p}, Pa$ & 67982 & 146334 & 179066 & 75879 & 116896 & 155675 \\ \hline
  $A_{tv}, m/s$ & 4.82 & 3.18 & 1.93 & 3.89 & 4.95 & 2.85 \\ \hline
  $\Delta \phi, ^{\circ}$ & 177 & 132 & 145 & 96 & 112 & 143 \\ \hline
\end{tabular}
\label{tab:2500at0dot33}
\end{center}
\end{table}

\subsection{The resonance in the injectors}
\label{sec:resoninj}

\begin{figure}[H]
    \begin{subfigmatrix}{2}
        \subfigure[pressure]
{\includegraphics{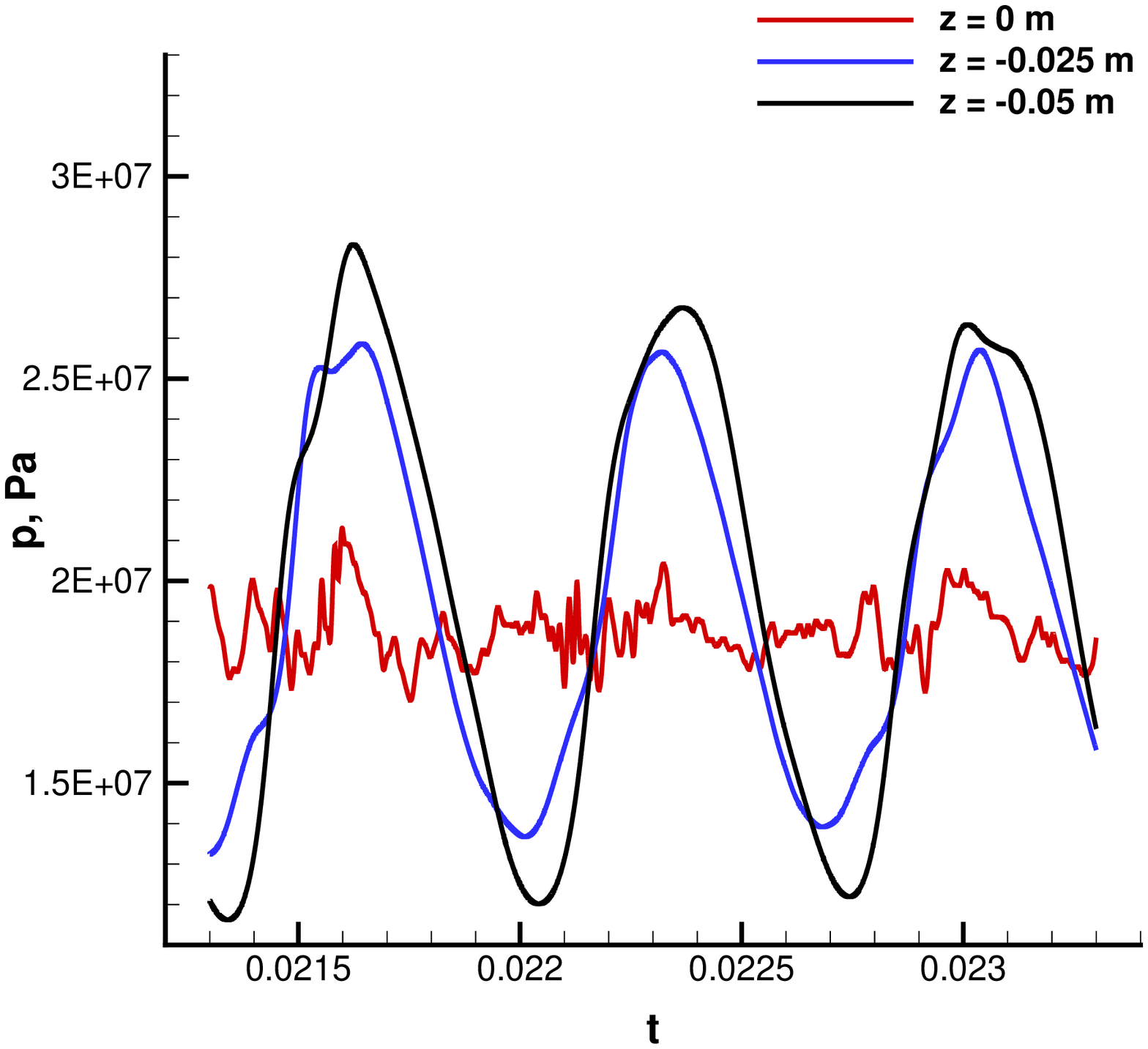}
        \label{fig:p_his_fpv_central_inj}}
        \subfigure[axial velocity]
{\includegraphics{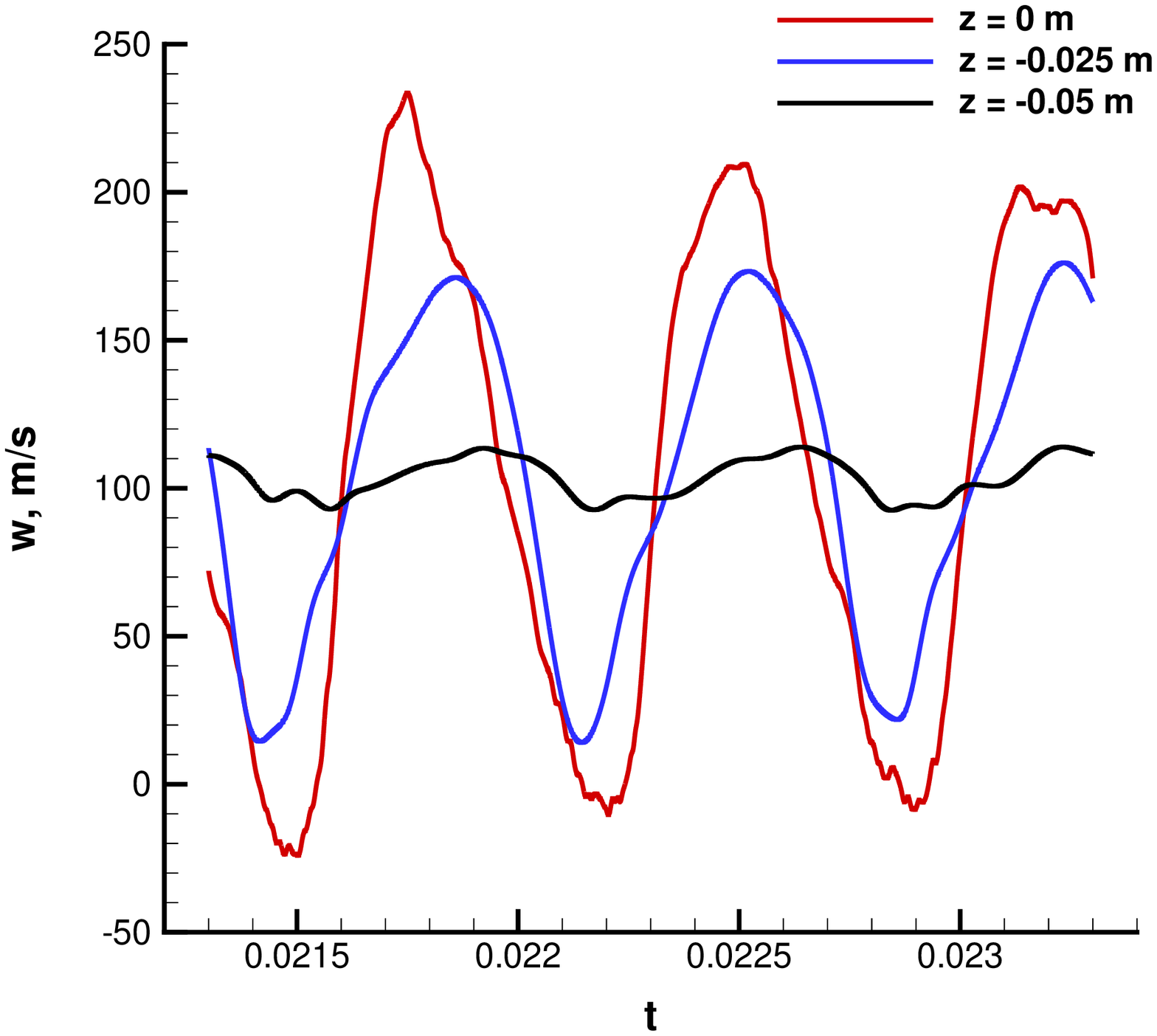}
        \label{fig:w_his_fpv_central_inj}}
    \end{subfigmatrix}
    \caption{Time histories of pressure and tangential velocity on the chamber axis inside the central oxidizer injector for the FPV approach}
    \label{fig:p_w_his_fpv_central_inj}
\end{figure}

\begin{figure}[H]
    \begin{subfigmatrix}{2}
        \subfigure[pressure]
{\includegraphics{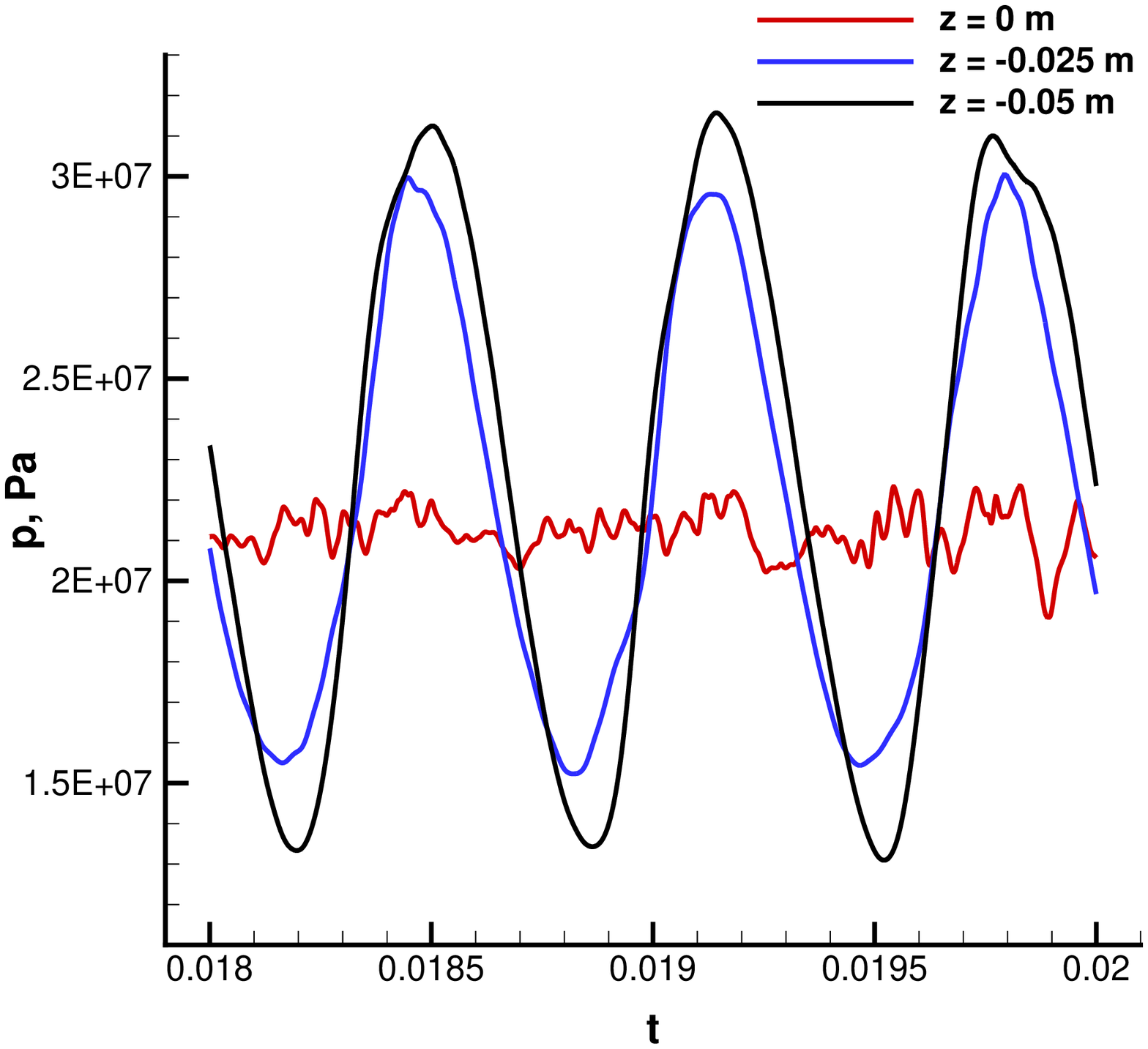}
        \label{fig:p_his_osk_central_inj}}
        \subfigure[axial velocity]
{\includegraphics{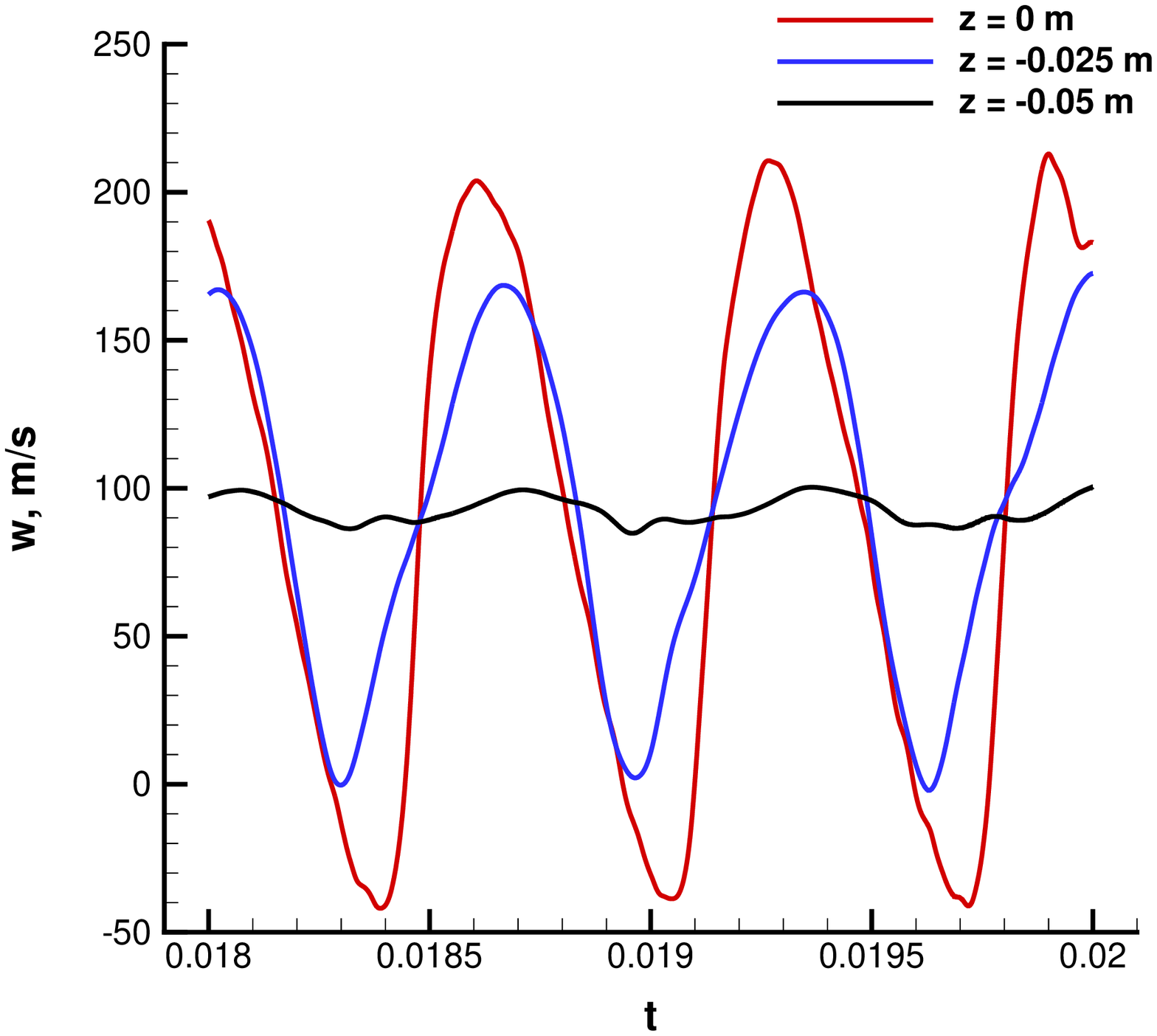}
        \label{fig:w_his_osk_central_inj}}
    \end{subfigmatrix}
    \caption{Time histories of pressure and tangential velocity on the chamber axis inside the central oxidizer injector for the OSK method}
    \label{fig:p_w_his_osk_central_inj}
\end{figure}

In this ten-injector analysis, the flow within the injector ports is 
simulated with a better resolution compared to that for configurations with
a larger number of injectors. The injector-plate face and injector exit
are at $z = 0$. The coordinate $z$ inside the injector is negative.
The injector port has a length of $0.05$ m with $z$
varying from $-0.05$ m to $0$ within the port. Figure
~\ref{fig:p_w_his_fpv_central_inj} 
shows the oscillatory behavior in the central oxidizer port from an analysis
using the FPV approach. Similar oscillatory behavior
occurs with the one-step kinetics model as shown in Figure
~\ref{fig:p_w_his_osk_central_inj}. This resonance phenomenon
is also found in each of the injectors. However, the oscillation
magnitude reaches a maximum value in the central injector with decreased
values in the surrounding injectors.

A pressure antinode (velocity node) occurs at $z = -0.05$ m, the
injector-port entrance, and a pressure node (velocity antinode) occurs
at $z = 0$, the injector exit.  At $z = -0.025$ m, the phase angle difference
between pressure and the axial velocity is about 90 degrees. All these
features suggests that the injector is behaving as a quarter-wave-length
tube. Here, the exit of the injector behaves as the open end while the
inlet is partially closed and behaves as the closed end. For pure oxygen
at 400 K, the speed of sound is about 379 m/s. The theoretical frequency
of this 5-cm quarter-wave-length resonance is 1895 Hz. The actual
resonance frequency here is 1500 Hz, deviating slightly from the
theoretical value. The oscillation at 1500 Hz inside the injector port is 
expected since it is
driven by the longitudinal-mode oscillation in the combustion chamber.

\section{Conclusion}

The combustion instability is computationally studied through the flamelet progress variable (FPV) approach for a ten-injector rocket engine. The governing equations for the FPV approach are solved by a newly developed C++ code which is based on OpenFOAM 4.0. The Flamemaster code is used to generate the flamelet tables for methane/oxygen combustion at a background pressure of $200$ bar. A 12-species kinetics is adopted to represent the chemical reactions. To address the pressure effect on the reaction rate for the progress variable while reducing computational effort in looking up the tables, the reaction rate is rescaled using a power law. The one-step-kinetics (OSK) method is also applied in computations as a comparison to the FPV apporach.

Fourier analysis of the computed time histories of pressure at the near-wall probes shows that the modes of $1500$ Hz and $2500$ Hz are dominant in both the FPV and OSK approaches. The amplitude of the $1500$ Hz mode for the FPV approach is equivalent to that for the OSK method. However, the amplitude of the $2500$ Hz mode is larger for the FPV approach. At the near-wall probes of six different azimuthal positions, pressure oscillations due to the $1500$ Hz mode are in phase especially near the nozzle entrance. At each probe, pressure is $90$ degrees out of phase with the axial velocity. Hence, the mode of $1500$ Hz is identified as a longitudinal mode. 
For the mode of $2500$ Hz, the time histories of pressure are $180$ degrees out of phase at any pair of the near-wall probes that are $180$ degrees away from each other. Examination of the mode amplitude for pressure and tangential velocity reveals that the azimuthal positions of the pressure node/antinode are the antinodal/nodal points of the tangential velocity. Therefore, the $2500$ Hz mode is essentially a tangential standing wave. However, possibly due to the impact of boundary layer on velocity phase, the phase angle difference between pressure and the tangential velocity is far from $90$ degrees for this mode.

Both the FPV and OSK approaches show that a resonance phenomenon occurs in the injectors. The oscillation magnitude achieves a maximum in the central injector while decreasing in the surrounding injectors. The injector outlet behaves as the open end, where a pressure node (velocity antinode) is observed. The injector inlet behaves as the close end and a pressure antinode (velocity node) occurs there. Pressure and the axial velocity are $90$ degrees out of phase. The resonance at $1500$ Hz in the injector port is anticipated as it is driven by the longitudinal mode of oscillations in the combustion chamber.

\section{Acknowledgments}

This research was supported by the U.S. Air Force Office of Scientific Research under grant FA9550-15-1-0033 , with Dr. Mitat Birkan as the program manager. Professor Heinz Pitsch of RWTH
Aachen University is acknowledged for providing us access to the FlameMaster code. Professor Hai Wang of Stanford University is acknowledged for providing us with the FFCMy-12 chemical reaction mechanism for the methane/oxygen combustion. We are also grateful to Dr. Tuan Nguyen of the Sandia National Laboratories for the insightful discussions.

\end{document}